\documentclass[journal=jacsat,manuscript=article,layout=twocolumn]{achemso}

\usepackage[version=4]{mhchem} 

\usepackage{color}
\usepackage{epstopdf}
\usepackage{graphicx}
\usepackage{subfig}
\usepackage{physics}
\usepackage{xcolor}
\usepackage[labelfont=bf]{caption}
\usepackage{bm}
\usepackage{amsmath, amsthm, amssymb, amsfonts}
\usepackage{mathtools}
\usepackage{multirow}
\usepackage{physics}
\usepackage{array}
\usepackage{dblfloatfix}
\usepackage{abstract} 
\usepackage[backref=none,bookmarksnumbered=true,bookmarks=true,bookmarksopen=true,colorlinks=true,citecolor=blue,linkcolor=blue,anchorcolor=green,urlcolor=blue,unicode=false]{hyperref}
\let\oldmaketitle\maketitle
\let\maketitle\relax

\usepackage{float}

\title{ Theory of scanning tunneling spectroscopy beyond one-electron molecular orbitals: can we image molecular orbitals?}

\author{Manish Kumar}
\affiliation{Institute of Physics, Czech Academy of Sciences, Prague, Czech Republic}
 \email{kumarm@fzu.cz}
 
\author{Diego Soler}
\affiliation{Institute of Physics, Czech Academy of Sciences, Prague, Czech Republic}
 \email{soler@fzu.cz}
 
\author{Marco Lozano}
\affiliation{Institute of Physics, Czech Academy of Sciences, Prague, Czech Republic}

\author{Enzo Monino}
\affiliation{Department of Theoretical Chemistry, J. Heyrovsky Institute of Physical Chemistry, Czech Academy of Sciences, Prague 18200, Czech Republic}

\author{Libor Veis}
\affiliation{Department of Theoretical Chemistry, J. Heyrovsky Institute of Physical Chemistry, Czech Academy of Sciences, Prague 18200, Czech Republic}

\author{Pavel Jelínek}
\affiliation{Institute of Physics, Czech Academy of Sciences, Prague, Czech Republic}
\affiliation{Czech Advanced Technology and Research Institute (CATRIN), Palacký University Olomouc, 779 00 Olomouc, Czech Republic}
 \email{jelinekp@fzu.cz}

\begin{document}

\twocolumn[
\begin{@twocolumnfalse}
\oldmaketitle
\begin{abstract}
\setstretch{0.9}
\noindent\hrulefill
\vspace{2mm}

The interpretation of experimental spatially resolved scanning tunneling spectroscopy (STS) maps of close-shell molecules on surfaces is usually interpreted within the framework of one-electron molecular orbitals. Although this standard practice often gives relatively good agreement with experimental data, it contradicts one of the basic assumptions of quantum mechanics, postulating the impossibility of direct observation of the wavefunction, i.e., individual molecular orbitals. The scanning probe community often considers this contradiction about observing molecular orbitals as a philosophical question rather than a genuine problem. Moreover, in the case of polyradical strongly correlated molecules, the interpretation of STS maps based on one-electron molecular orbitals often fails. Thus, for a precise interpretation of STS maps and their connection to the electronic structure of molecules, a theoretical description, including non-equilibrium tunneling processes going beyond one-electron process, is required.

In this contribution, we first show why, in selected cases, it is possible to achieve good agreement with experimental data based on one-electron canonical molecular orbitals and Tersoff-Hamann approximation. Next, we will show that for an accurate interpretation of strongly correlated molecules, it is necessary to describe the process of removing/adding an electron within the formalism of many-electron wavefunctions for the neutral and charged states. This can be accomplished by the concept of so-called Dyson orbitals. We will discuss the examples where the concept of Dyson orbitals is mandatory to reproduce experimental STS maps of polyradical molecules. Finally, we critically review the possibility of the experimental verification of the so-called SOMO/HOMO inversion effect using STS maps in polyradical molecules. Namely, we will demonstrate that experimental STS measurements cannot provide any information in case of strongly correlated molecules about the ordering of one-electron molecular orbitals and, therefore neither about the SOMO/HOMO inversion effect.

\noindent\hrulefill

 \end{abstract}
\end{@twocolumnfalse}
]

\setstretch{0.9}
 
\clearpage

\setstretch{0.9}

\section{Introduction}
Scanning tunneling microscopy (STM) \cite{Binnig1982} is a unique tool that allows us not only to image objects with atomic resolution, but also to obtain valuable information about the electronic structure \cite{Chen2007}. Namely, scanning tunneling spectroscopy (STS) \cite{Stroscio1986,Zandvliet2009} provides information about the electronic structure of individual molecules \cite{Chiang1997} on the surfaces of solids. The spatial dI/dV maps obtained by STS measurements are often interpreted using one-electron frontier molecular orbitals \cite{repp2005molecules}. Moreover, in many cases, dI/dV maps are associated directly with images of molecular orbitals, due to the coincidence of their spatial distribution. However, this interpretation is in sharp contradiction with one of the basic postulates of quantum mechanics, which excludes direct observation of the wavefunction of a quantum system. Therefore, this approach received a lot of criticism, especially from the chemistry community, see e.g. \cite{pham2017can,Krylov2020,Scerri2000}. However, this dichotomy is very often defended in the STM community, due to the relatively good agreement between experimental data and calculated one-electron molecular orbitals, as a {\it philosophical} question. However, the interpretation of STS data based on one-electron molecular orbitals can be quite problematic, especially in the case of strongly correlated molecules. For example, the recent development of on-surface synthesis \cite{Zhao2022,Li2020,Pavliek2017,su2023atomically,Song2024,zuzak2024surface,Villalobos2025,calupitan2023emergence,vilas2023surface} has enabled the preparation of highly radical molecules with strongly correlated electronic structures. As practice shows, it is precisely in these cases that the interpretation based on one-electron molecular orbitals often fails \cite{Villalobos2025,Toroz2013,Schulz2015}. Moreover, recently, molecules exhibiting an~unusual arrangements of molecular orbitals, where the energy level of the HOMO orbital is higher than that of the SOMO orbital, have received considerable attention \cite{Kasemthaveechok2022,Kumar2017,Westcott2000,Kusamoto2008}. This effect, called SOMO/HOMO inversion (SHI), has recently been discussed in connection with STM measurements of diradical molecules\cite{Mishra2024}.

We should note that the problem of the correct interpretation of dI/dV maps discussed here was already addressed by several groups before \cite{Toroz2013,yu2017apparent,Schulz2015}. But we feel that a concise picture accessible also for general audience, including a simple guideline for the interpretation, is still missing. Here we introduce concept of multi-reference Dyson orbitals \cite{truhlar2019orbitals,ortiz2020dyson}, which enable us to describe the non-equilibrium process during the single electron tunneling through the molecules simultaneously considering strongly correlated electronic structure of the polyradical molecules.

The article is organized as follows. First, we briefly recapitulate the concept of one-electron molecular orbitals. We summarize the one-electron theory describing tunneling in STM junction in quasi-equilibrium conditions and we show the connection with one-electron molecular orbitals.
Next, we then introduce the concept of Dyson orbitals, which importance we demonstrate on selected cases of strongly correlated molecules. Namely, we make a comparison with available experimental STS measurements. Finally, we critically discuss the possibility of detecting the SHI effect using STS measurements.

\section{Results and Discussion}
\subsection{The concept of molecular orbitals}

For the sake of simplicity we briefly summarize the concept of one-electron molecular orbitals (MO), which plays a fundamental role in our understanding of the physical and chemical properties of molecular compounds. They represent a spatial localization of individual electrons in molecule, but they are not uniquely defined quantities. Strictly speaking, MOs are mathematical objects which represent eigenfunctions of a given one-electron quantum operator.  
A complete description of an electron must include its spin, therefore, the fundamental one-electron functions correspond to molecular spin-orbitals. When referring only to their spatial component, they are referred to as MOs.
In the case of canonical Kohn-Sham (KS) spin orbitals, $\phi_i^{KS}(\mathbf{x})$, they are obtained in the framework of density functional theory (DFT) solving the set of one-electron Kohn-Sham equations of the $N$-electron system \cite{Kohn1965}:
\begin{equation}
\label{eq: one electron hamil}
  \left( -\dfrac{\hbar}{2m} \nabla^2 + V_{KS}(\mathbf{x})    \right) \phi^{KS}_{i}(\mathbf{x}) = \epsilon_{i}\phi^{KS}_{i}(\mathbf{x}), 
\end{equation}
where, $V_{KS}$ represents an~effective external potential acting on $i-th$ non-interacting electron, where $i=1,..,N$. The variable $\mathbf{x}=\left( \mathbf{r}, \sigma \right)$ labels both position $\mathbf{r}$ and spin $\sigma $.  Variable $\epsilon_{i}$ represents the spin orbital energy of the corresponding one-electron KS canonical spin orbital $\phi^{KS}_{i}$. In the DFT framework, the many-electron ground state $\Psi^{KS} \left(\mathbf{x}_{1},\cdots,\mathbf{x}_{N}\right)$ of the $N$-electron system is described as a single determinant of the $N$ lowest KS spin orbitals $\phi^{KS}_{i}(\mathbf{x})$:  
\begin{align}
\label{eq: definiton of Slater determinant}
    \Psi^{KS} & \left(\mathbf{x}_{1},\cdots,\mathbf{x}_{N}\right) =  
    \frac{1}{\sqrt{N!}} \nonumber \\
    & \times
    \begin{vmatrix} 
        \phi^{KS}_1(\mathbf{x}_1) & \phi^{KS}_2(\mathbf{x}_1) & \cdots & \phi^{KS}_N(\mathbf{x}_1) \\
        \phi^{KS}_1(\mathbf{x}_2) & \phi^{KS}_2(\mathbf{x}_2) & \cdots & \phi^{KS}_N(\mathbf{x}_2) \\
        \vdots & \vdots & \ddots & \vdots \\
        \phi^{KS}_1(\mathbf{x}_N) & \phi^{KS}_2(\mathbf{x}_N) & \cdots & \phi^{KS}_N(\mathbf{x}_N)
    \end{vmatrix}.
\end{align}
Note that the KS many-electron ground state $\Psi^{KS}$ is a single-reference state, as it is described by a single Slater determinant (eq. \ref{eq: definiton of Slater determinant}). Therefore, it fails to describe strongly correlated open-shell systems 
as will be discussed later on. To describe correctly strongly correlated open-shell systems of $N$-electrons, we have to used multi-reference many-electron  wavefunction $\Psi \left(\mathbf{x}_{1},\cdots,\mathbf{x}_{N}\right)$, which consist of linear combinations of several Slater determinants with non-negligible contributions to the ground state.  

For clarity, throughout this article, we employ small greek letters ($\phi,\varphi,\chi$...) to denote single-particle wavefunctions (i.e., orbitals) and capital greek letters (i.e., $\Psi$) to denote many-electron states.  

Natural orbitals $\phi^{NO}_{i}(\mathbf{r})$ are another type of MOs, which 
are known to provide the most compact representation of correlated wavefunctions \cite{szabo1989modern}.
The set of $N$ natural orbitals are eigenvectors obtained from diagonalization of the spinless 1-particle reduced density matrix $\gamma(\mathbf{r},\mathbf{r}^\prime)$ :
\begin{equation}
    \label{eq: diagonalization of rho}
    \int \gamma(\mathbf{r},\mathbf{r}^\prime) \phi_j^{NO}(\mathbf{r}^\prime) d\mathbf{r}^\prime = \lambda_j\phi_j^{NO}(\mathbf{r}),
\end{equation}
where  the  density matrix $\gamma(\mathbf{r},\mathbf{r}^\prime)$ is defined in the position representation: 
\begin{align}
    \gamma(\mathbf{r},\mathbf{r}^\prime) &= N \sum_{\sigma, \sigma^\prime} \int  
    \Psi^\ast \left( \mathbf{r} \sigma, \mathbf{x}_{1},\cdots,\mathbf{x}_{N-1} \right) \nonumber \\
    &\quad \times \Psi \left( \mathbf{r}^\prime \sigma^\prime, \mathbf{x}_{1},\cdots,\mathbf{x}_{N-1} \right) \nonumber \\
    &\quad \times d\mathbf{x}_{1} \cdots d\mathbf{x}_{N-1}.
    \label{eq: density matrix}
\end{align}
The positive eigenvalues $\lambda_j$ in eq. \ref{eq: diagonalization of rho} represent the occupations of the natural orbitals $\phi_j^{NO}$. The occupied and unoccupied  natural orbitals have the occupancy $\lambda_j$ close to 2 or 0, respectively. The natural orbitals $\phi_j^{NO}$ whose occupations have fractional values significantly different from integer values of 2 or 0 contribute to the number of unpaired electrons, with $\lambda=1$ means one whole unpaired electron (i.e., greatest radical character for the given orbital). 

Unitary transformations can convert delocalized canonical orbitals into localized molecular orbitals (LMOs) and vice versa. LMOs help bridge molecular Lewis structures with quantum-chemical calculations, facilitate the interpretation of correlations between molecular fragments \cite{Song2024}, and enable a more efficient treatment of dynamic electron correlation \cite{Eriksen2015}.

It should be stressed that the one-electron MOs are experimentally not observable.  According to the basic postulates of Quantum Mechanics, the \textit{observables} of a system are self-adjoint operators, $A$, acting on physical states $\Psi$.  The \textit{measurable} quantities are the eigenvalues of such operators represented in a~given basis set. 
If a many-electron wavefunction $\Psi$ describing a given system has single-reference character, i.e. can be approximated by a single Slater determinant of a set of MOs, $\{ \phi \}$, the expected value of observable $A$, denoted by $\langle A \rangle$, can be expressed as
\begin{equation}
    \label{eq: expected value one electron}
    \langle A \rangle = \sum_j \langle \phi_j \vert A \vert \phi_j \rangle.
\end{equation}
This explains why one-electron MOs $\phi$ can be used as tools to compute experimentally observable quantities. However, as such, the one-electron MOs $\phi$ themselves are purely mathematical objects that do not have any physical meaning. Hence MOs are not directly observable, which contradicts the general practice where $dI/dV$ measurements obtained from STM are interpreted on the basis of one-electron MOs.

\subsection{Bardeen Perturbation theory of tunneling}
\begin{figure*}[htbp]
    \centering
    \includegraphics[width=\textwidth]{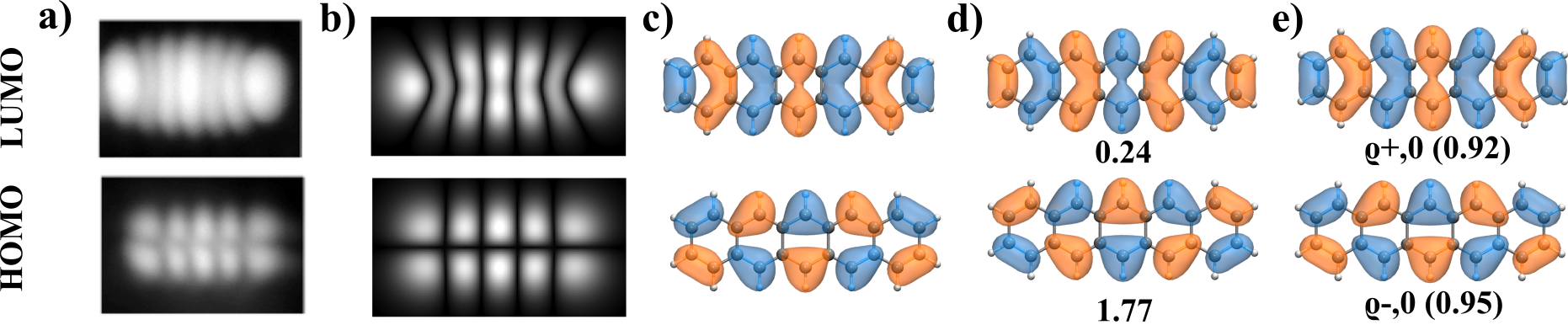}
    \caption{a) Experimental STM maps adopted from Ref.\cite{repp2005molecules}; b) simulated dI/dV maps of canonical DFT orbitals obtained from spin-unpolarized DFT-PBE0 calculations using PP-STM code with metal-like tip \cite{Krej2017}; c) canonical DFT orbitals obtained from spin-unpolarized DFT-PBE0 calculations; d) multi-reference natural orbitals obtained from CASCI(12,12) calculations with corresponding occupation;  e) multi-reference Dyson orbitals obtained from CASCI(12,12) calculations with corresponding strengths for the pentacene molecule.}
    \label{fig: pentacene-orb}
\end{figure*}

The generally adopted theory of scanning probe microscopy to describe the tunneling process between tip and sample is based on the time-dependent first-order perturbation theory \cite{Chen2007}. For detailed derivation of the tunneling theory, we refer readers to the excellent book of C.J. Chen \cite{Chen2007}. The theory assumes the resonant tunneling process between tip and sample both represented by the ground state wavefunctions. Moreover, it also considers the quasi-equilibrium situation, where the applied bias voltage $eV$ between tip and sample causes only a rigid shift of the density of states of the sample $\rho_{s}(E_F-eV)$ with respect to the density of states of the tip $\rho_{t}(E_F)$. The conductance $G=\cfrac{dI}{dV}$ at a particular applied bias voltage $eV$ between tip and sample can be expressed as follows:  
\begin{equation}
    \label{eq: current_Bardeen}
   \frac{dI}{dV} = \frac{4\pi e}{\hbar} \rho_{s}(E_F-eV)\rho_{t}(E_F)\vert M_{ts} \vert^2,
\end{equation}
$E_F$ means the Fermi energy. The transition matrix between tip and sample $M_{ts}$ is expressed by Bardeen's surface integral:
\begin{equation}
    \label{eq: matrix_Bardeen}
   M_{ts} = -\frac{\hbar^2}{2m} \int_{S}\left( \phi^*_t\nabla\phi_s - \nabla \phi_t \phi^*_s    \right) dS,
\end{equation}
where $\phi_{t,s}$ represents one-electron wavefunctions of tip and sample, respectively, and $m$ is the mass of the electron. The matrix element $M_{ts}$ is defined by the selection rules imposed by the symmetry of the tip $\phi_t$ and sample $\phi_s$ wavefunctions.

In the so-called Tersoff-Hamann approach, eq. \ref{eq: current_Bardeen} can be further simplified, provided that the tip wavefunction $\phi_t$ can be represented by a single s-like orbital and the constant density of state of tip $\rho_{t}(E) \approx const$ \cite{Tersoff1985}. Thus, the tunneling current is proportional to the local density of the states of the surface $\rho_s$, and eq. \ref{eq: current_Bardeen} can be simplified as follows: 

\begin{equation}
    \label{eq: current_TH}
   \frac{dI}{dV} \approx \rho_{s}(\epsilon-eV).
\end{equation}
In the case of physisorbed molecules on metal surfaces with negligible hybridization and charge transfer between the molecule and surface, the density of states $\rho_{s}$ can be approximated by the set of canonical or natural one-electron molecular orbitals typically obtained by solving one-electron Kohn-Sham equations \cite{Kohn1965}. Thus, for particular bias voltages, the density of states $\rho_{s}$ reduces to a particular canonical orbital, which is in resonance with the Fermi level of the tip; i.e. $\rho_{s}=\vert \phi_i \vert^2$. Consequently, dI/dV measurements within the Tersoff-Hamann approach can be directly associated with a particular molecular orbital $\phi_i$.  This is a remarkable conclusion, which is often used by SPM community to justify the interpretation of the experimental dI/dV maps within the framework of one-electron molecular orbitals. 

Indeed, the long-standing experience shows that such association with molecular orbitals can often provide a reasonable agreement with the experimental evidence. For example, in the seminal work of Repp et al. \cite{repp2005molecules}, they showed that spatial STM maps acquired at given low bias voltages on pentacene molecules with a metallic tip (see Figure \ref{fig: pentacene-orb}a) exhibit striking agreement with the one-electron canonical highest occupied molecular orbital (HOMO) and lowest unoccupied molecular orbital (LUMO) obtained from DFT (see Figure \ref{fig: pentacene-orb}c) as well as their corresponding dI/dV maps simulated with tip represented by s-like orbital (see Figure \ref{fig: pentacene-orb}b).

According to the multi-reference complete active space configuration interaction (CASCI) calculations employing the active space of 12 electrons in 12 orbitals, CASCI(12,12), the ground state wavefunction $\Psi_o$ of the neutral pentacene molecule has a dominant contribution of a single Slater determinant with a doubly occupied HOMO orbital, see middle Figure \ref{fig: pentacene-spectral_func}a). Consequently, the occupancy of the natural frontier HONO and LUNO orbitals is 1.77 and 0.24, respectively, (see in Figure \ref{fig: pentacene-orb}d),  indicating the close-shell character of the pentacene molecule. Comparison of Figures  \ref{fig: pentacene-orb} c) and d) reveals a very similar character of one-electron canonical DFT orbitals and natural orbitals obtained from multi-reference CASCI(12,12) calculations. 

This example demonstrates that the dI/dV maps of weakly bounded molecules on surfaces with a strong closed shell character can often be reasonably well interpreted on the basis of one-electron MOs. This experience led to a common practice, which associates the spatial distribution of dI/dV maps of weakly bounded molecules on surfaces with frontier one-electron MOs using the Tersoff-Hamann approach.  In this context, the criticism of this approach could be viewed as a simple philosophical discussion, which does not have any implications for the correctness of the interpretations using the association with the concept of one-electron MOs. However, as frequently happens, evil resides in some details, as will be discussed later on. 

\subsection{Concept of Dyson orbitals}
\begin{figure*}[ht]
    \centering
    \includegraphics[width=1\linewidth]{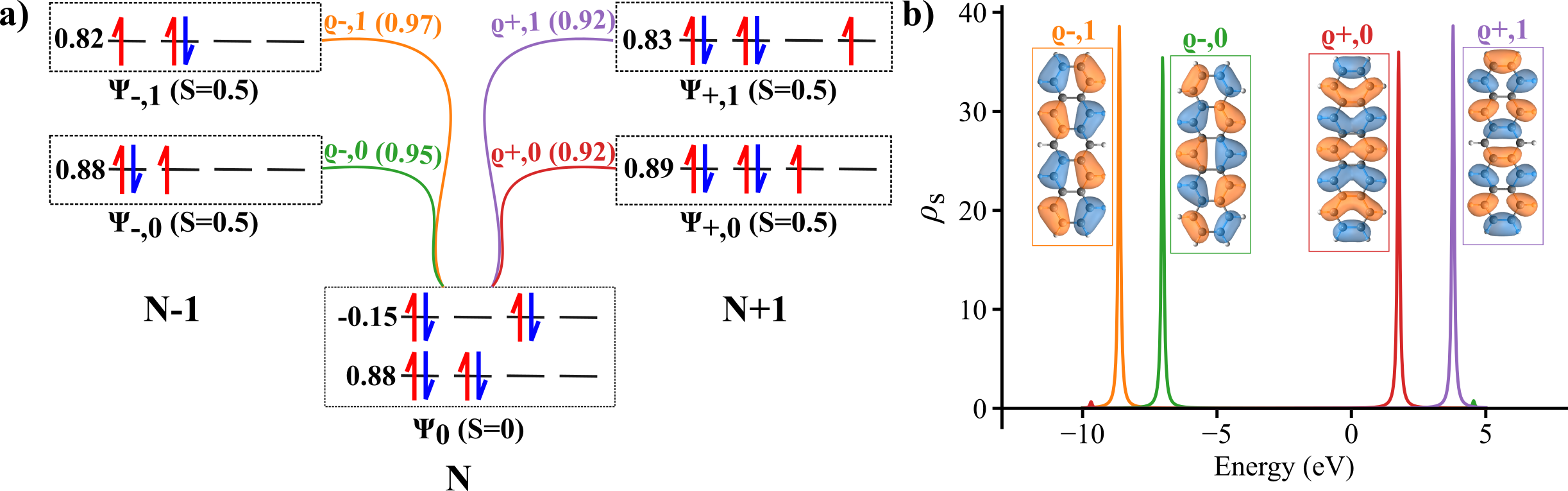}
    \caption{ a) Multi-reference wavefunctions of neutral ground state and charge state multiplet. Horizontal lines represent individual molecular orbitals (MOs) solely for occupancy purposes and do not indicate degeneracy. This format is consistently used throughout the article. b) Multi-reference spectral function for the pentacene molecule.}
    \label{fig: pentacene-spectral_func}
\end{figure*}




The STM perturbation theory sketched in the previous section is limited to a one-electron formalism at quasi-equilibrium conditions.  Therefore, the perturbation formalism has two drawbacks. First, it cannot capture fully the tunneling process through strongly correlated molecular systems, where the one-electron picture fails. In the case of open-shell strongly correlated molecules,  we employed multi-reference CAS methods to describe correctly their electronic structure. Second, it does not explicitly consider a~temporal addition/removal of an electron in the molecule during the tunneling process between tip and sample. In other words, the change of electron density associated with the transition from the ground state of the neutral molecule $\Psi^N$ to the charged states $\Psi^{N\pm 1}$ caused by the tunneling of electrons through the molecule during the measurement are completely neglected. 

To describe the transition from the neutral ground state to low-energy charged (N$\pm1$) states of strongly correlated molecules, we use the concept of so-called Dyson orbitals \cite{ortiz2020dyson}. We consider multi-reference wavefunctions of the ground state of the neutral molecule $\Psi_0(\mathbf{x}_1,\cdots,\mathbf{x}_N)$, and a manifold of low-energy states for the negatively and positively charged system $\Psi_{+,j}(\mathbf{x}_1,\cdots,\mathbf{x}_N,\mathbf{x}_{N+1})$,  $\Psi_{-,j}(\mathbf{x}_1,\cdots,\mathbf{x}_{N-1})$. Dyson spin orbitals are then defined as the overlaps between the neutral ground state $\Psi_0$ and $j$-th charged states $\Psi_{\pm,j}$:
\begin{align}
    \varrho_{+,j}(\mathbf{x}) &= \langle \Psi_{+,j} \vert \hat{\psi}^\dagger(\mathbf{x}) \vert \Psi_0  \rangle \nonumber \\
    &= \sqrt{N+1}  \int \Psi_0(\mathbf{x}_1,\cdots,\mathbf{x}_N)^\ast \nonumber \\
    &\quad \times \Psi_{+,j}(\mathbf{x},\mathbf{x}_1,\cdots,\mathbf{x}_{N})  
    d\mathbf{x}_1 \cdots d\mathbf{x}_{N}, \nonumber \\
    \varrho_{-,j}(\mathbf{x}) &= \langle \Psi_{-,j} \vert \hat{\psi}(\mathbf{x}) \vert \Psi_0  \rangle \nonumber \\
    &= \sqrt{N} \int \Psi_0(\mathbf{x},\mathbf{x}_1,\cdots,\mathbf{x}_{N-1})^\ast \nonumber \\
    &\quad \times \Psi_{-,j}(\mathbf{x}_1,\cdots,\mathbf{x}_{N-1})  
    d\mathbf{x}_1 \cdots d\mathbf{x}_{N-1}.
    \label{eq: definition Dyson orbitals}
\end{align}

Here $\hat{\psi}^\dagger(\mathbf{x})$ and $\hat{\psi}(\mathbf{x})$ are the field operators for the creation and annihilation of an electron at position $(\mathbf{x})$.
Importantly, Dyson orbitals represent the transition amplitudes from the neutral ground state to a particular charged state $N\pm1$. Therefore, we can construct the multi-reference spectral function $\rho_s$, which represents  the local density of states of the molecular system upon charging by a single electron as follows: 
\begin{align}
    \rho_s(\mathbf{x},\omega) &= \eta \sum_j 
    \frac{\vert \langle \Psi_{+,j} \vert \hat{\psi}^\dagger(\mathbf{x}) \vert \Psi_0  \rangle \vert^2}
    {\left( \omega - E_j  \right)^2 + \eta^2} \nonumber \\
    &\quad + \eta \sum_j 
    \frac{\vert \langle \Psi_{-,j} \vert \hat{\psi}(\mathbf{x}) \vert \Psi_0  \rangle \vert^2}
    {\left( \omega - E_j  \right)^2 + \eta^2},
    \label{eq: DOS Lehmann}
\end{align}
where $E_j$ is the energy difference of the charged states $\vert \Psi_{\pm,j} \rangle$ with respect to the neutral ground state, $\vert \Psi_0  \rangle$, $\omega$ is the energy and $\eta$ is the broadening of the resonance peaks, linked to a finite life-time of the excitations. Thus, the  density of state $\rho_s(\mathbf{x},\omega)$ is directly related to the Dyson orbitals. By substituting eq. \ref{eq: DOS Lehmann} defining the multi-reference density of states into the Tersoff-Hamann formula given by eq. \ref{eq: current_TH}, we see that the spatial distribution of the dI/dV maps is directly determined by the Dyson orbitals. From this point of view, Dyson orbitals are superior to single-electron MOs, because they include the transition probabilities between neutral and charged multi-reference states properly describing strongly correlated molecular systems. So the Dyson orbitals are observables because they remain invariant of the chosen basis set in which they are constructed. This can be demonstrated in the case of the benzene molecule as discussed in detail in supplementary material.

In the case of single-reference close shell molecules, it can be shown that Dyson orbitals reduce to single-electron canonical frontier orbitals \cite{ortiz2020dyson}. If we employ the one-electron molecular orbital picture, the wavefunctions are described by a single Slater determinant and these integrals reduce to one-particle integrals. If the basis of orbitals is the same for neutral and charged systems, then $\varrho_{+,j},\varrho_{-,j}$ coincide exactly with such one-electron canonical orbitals (see the supplementary material for a detailed proof).

For clarity, let us discuss the case of the pentacene molecule, to demonstrate the capability of the Dyson orbitals to describe the spatial dI/dV maps of close-shell molecules. According to multi-reference CASCI(12,12) calculation performed using the DFT orbitals shown in Figure S2, the ground state wavefunction $\Psi_0$ of pentacene is predominantly represented by a single Slater determinant with a doubly occupied HOMO orbital, as shown in Figure \ref{fig: pentacene-spectral_func}a). The occupancies of highest occupied natural orbital (HONO) and lowest unoccupied natural orbital (LUNO) shown in Figure \ref{fig: pentacene-orb}d), confirm the close-shell character of the pentacene molecule.

 According to the CAS calculations, the ground $\Psi_{0,\pm}$ and first excited $\Psi_{1,\pm}$ states of charged (N$\pm$1) systems, displayed in Figure \ref{fig: pentacene-spectral_func}a), are well-described by a single-reference state as well. Consequently, the electron (de)attachment during the tunneling process takes place exclusively in one MO, as shown schematically in Figure \ref{fig: pentacene-spectral_func}a).
 
Figure \ref{fig: pentacene-spectral_func}b) shows the multi-reference spectral function $\rho_s(\omega)$ of the pentacene molecule including the corresponding lowest energy Dyson orbitals for the first  $\varrho_{\pm,0}$  and second $\varrho_{\pm,1}$ positive/negative ionic resonances. Comparing Figures \ref{fig: pentacene-orb}c-e), we see that the Dyson orbitals $\varrho_{\pm,0}$ for the first ionic resonances are almost identical with one-electron HOMO/LUMO orbitals. This observation can be rationalized by the aforementioned fact that the ground $\Psi_{\pm,0}$ and first excited $\Psi_{\pm,1}$ states of charged states corresponding to the (de)attachment of an electron from the ground state $\Psi_0$ of the neutral molecule can be described well by a single Slater determinant, see Figure \ref{fig: pentacene-spectral_func}a). Namely, the ground state $\Psi_{+,0}$ of charge state $N+1$ is represented by the Slater determinant with an extra electron in LUMO, while the ground state $\Psi_{-,0}$ of $N-1$ state by the Slater determinant with a missing electron in HOMO. Consequently, the single electron charging processes take place dominantly through HOMO/LUMO orbitals. This fact explains why single electron HOMO/LUMO orbitals may reproduce well dI/dV maps of pentacene molecule at the first ionization processes in both bias polarities. 

\subsection{ Polyradical Open-Shell Molecules and Dyson Orbitals}
\begin{figure*}[t!]
    \centering
    \includegraphics[width=1\linewidth]{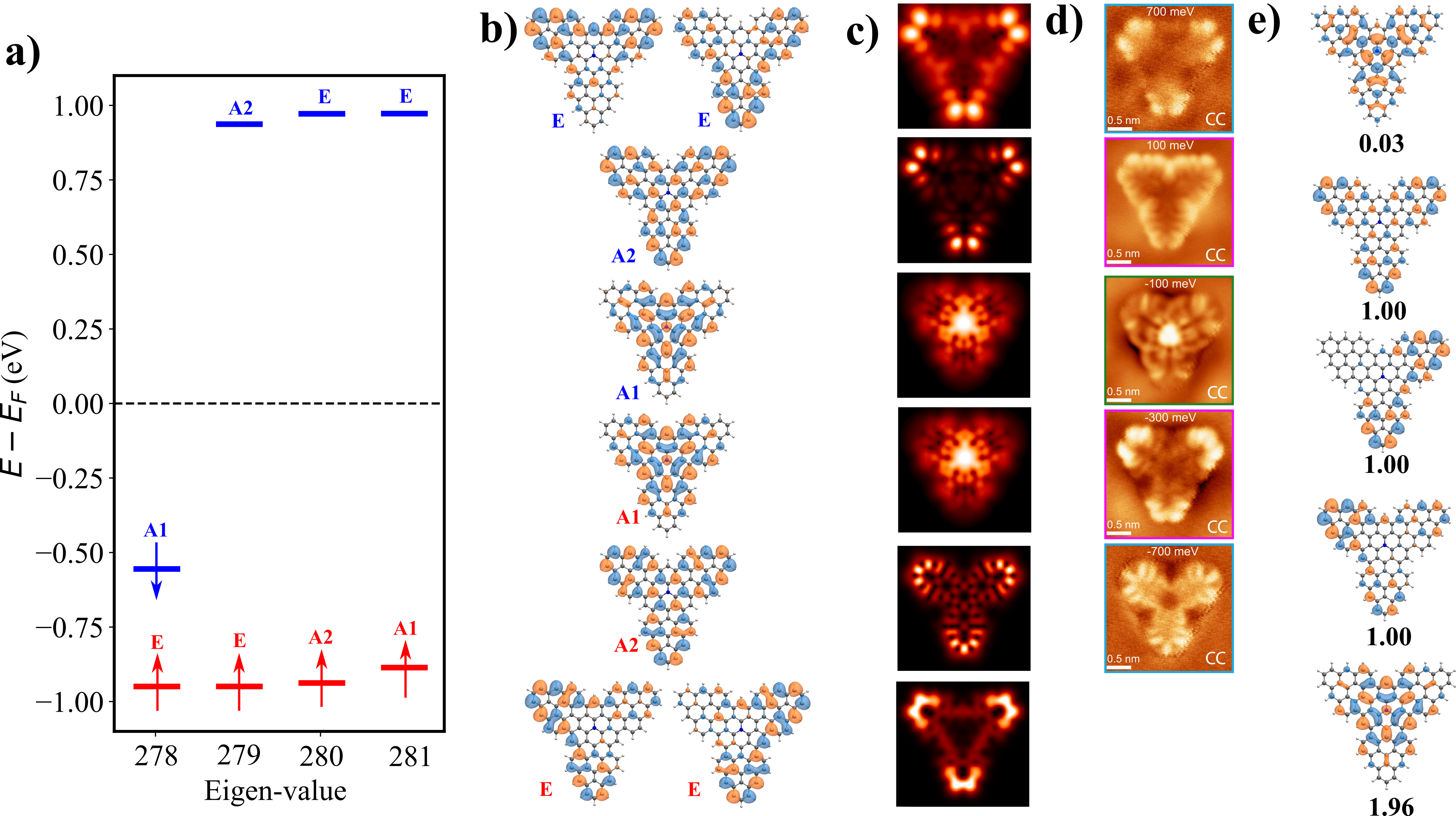}
    \caption{a) Spin polarized DFT-PBE0 eigen-value; b) canonical DFT orbitals obtained from spin polarized DFT-PBE0 calculations; c) simulated dI/dV maps of corresponding DFT-PBE0 orbitals; d) experimental dI/dV maps adapted with permission from ref. \cite{Vegliante25} Copyright 2025 American Chemical Society; e) multi-reference natural orbitals obtained from CASCI(11,11) calculation with corresponding occupation for \textbf{TTAT} molecule.}
    \label{fig: TTATA_dft}
\end{figure*}
Next, we will discuss two examples of strongly correlated polyradical molecules where the concept of Dyson orbitals obtained from multi-reference wavefunctions is mandatory to explain available experimental dI/dV measurements. 

We begin with polyradical open-shell triradical aza-triangulene (\textbf{TTAT}) molecule, synthesized by Vegliante et al \cite{Vegliante25} on an Au(111) surface, which possesses a high-spin quartet (S=3/2) ground state. Figures \ref{fig: TTATA_dft}a-b) illustrate calculated energy spectrum and corresponding canonical MOs obtained from spin-polarized DFT calculation using PBE0 exchange-correlation functional \cite{Adamo1999_PBE0}. Here blue/red color denotes states in down/up spin channels, respectively. 
According to this one-electron MO diagram, the first and second electron detachment, corresponding to negative ionic resonances (NIR), take place at the states labeled A1, which are energetically distinct corresponding to the opposite spin channel. The subsequent electron detachments occur from A2 MO followed by doubly degenerate E MOs. On the other hand, the first and second electron attachment, related to positive ionic resonances (PIR), are represented by A2 and E MOs, respectively.

\begin{figure*}[ht]
    \centering
    \includegraphics[width=1\linewidth]{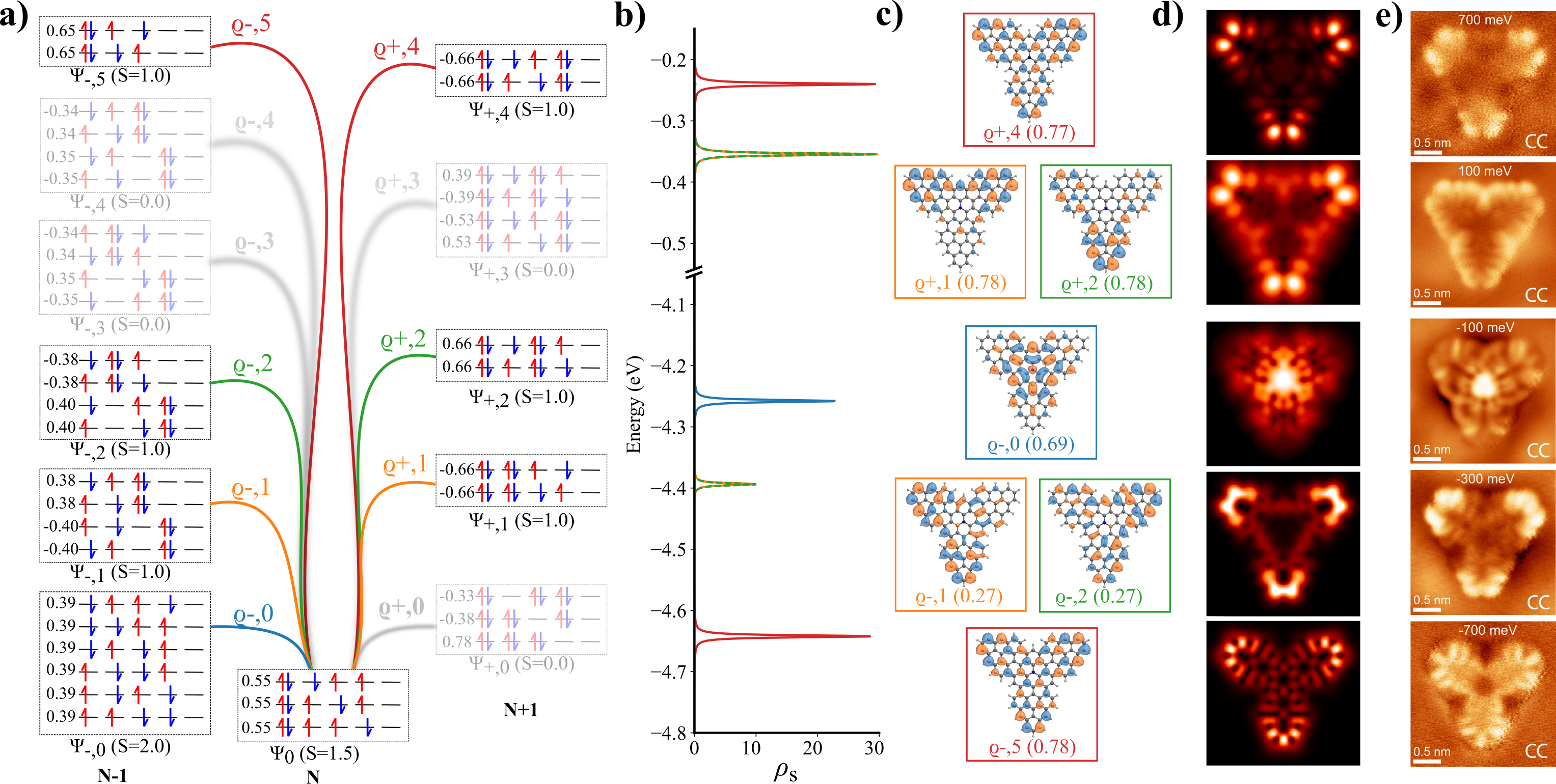}
    \caption{ a) Multi-reference wavefunctions of neutral ground state and charge state multiplets; b) calculated multi-reference spectral function;  c) multi-reference Dyson orbitals obtained from CASCI(11,11) calculations with corresponding strengths; d) simulated dI/dV of corresponding Dyson orbitals using PP-STM code \cite{Krej2017} with CO-like tip; e)  experimental dI/dV maps adapted with permission from ref. \cite{Vegliante25} Copyright 2025 American Chemical Society for \textbf{TTAT} molecule.  }
    \label{fig: TTATA_Dyson}
\end{figure*}

Figure \ref{fig: TTATA_dft}c) displays simulated dI/dV maps corresponding to canonical MOs shown in Figure \ref{fig: TTATA_dft}b). If we compare them with the experimental dI/dV maps acquired for different ionic resonances (IR) shown in Figure \ref{fig: TTATA_dft}d), we find poor agreement between them. Only A1 SOMO is able to match with the experimental dI/dV map of the first NIR acquired at -100 meV. However, the second A1 orbital is completely missing. Also, in the case of doubly degenerate E and A2 SOMO states, their appearance in dI/dV maps seems to be exchanged.   
Similarly, the experimental dI/dV map of the first PIR acquired at 100 meV does not fit with the calculated dI/dV maps corresponding to A2 LUMO at all. Instead, the first PIR matches better two doubly degenerate E unoccupied orbitals. These discrepancies show that the one-electron canonical MOs cannot explain sufficiently the dI/dV maps of   \textbf{TTAT} molecule. 

We carried out CAS calculations using the orbitals from restricted open-shell Kohn-Sham shown in Figure S3 to obtain multi-reference electronic states for neutral and charged \textbf{TTAT} molecule, which reveal the quartet ground state (S=3/2) of neutral \textbf{TTAT} with three unpaired electrons according to the occupancies of NOs, shown in Figure \ref{fig: TTATA_dft}e). Due to the presence of three unpaired electrons in the neutral state and four unpaired electrons in the negatively charged state, the wavefunction exhibits strong multi-reference character, as illustrated in Figure \ref{fig: TTATA_Dyson}a). The ground-state wavefunction for the neutral molecule in the  $S_z = 1.5$  subspace can be approximated by a single Slater determinant, as shown in Figure S6. However, for the negatively charged state, the presence of four unpaired electrons requires a multi-reference wavefunction for an accurate description.

There is a question, whether we can use multi-reference natural orbitals instead of Dyson orbitals to rationalize the experimental STM/STS maps. First, we should note that multi-reference NOs do not take into account transition probabilities from the neutral ground state $\Psi_0$ to charged states $\Psi_{\pm,j}$, as Dyson orbitals do. In the next, we will see that these transitions may have serious implications on the STS spectra.
Second, natural orbitals do not possess energy ordering, as they are categorized by occupation numbers instead of energy levels.  This inherent trait presents difficulties for employing natural orbitals in direct comparisons with experimental data, particularly in systems where precise energy-level alignment is imperative. As depicted in Figure \ref{fig: TTATA_dft}e), there are three natural orbitals with an occupation of 1 that should appear in the first two ionization sequences. However, Figure \ref{fig: TTATA_dft}d) reveals that the initial ionization arises from one of the doubly natural orbitals. On the other hand, the ionization event at -100 meV is attributable to a blend of many SONO orbitals. From the discussion above, we see that multi-reference natural orbitals cannot be regarded as a substitute for Dyson orbitals when comparing dI/dV spectra with experimental maps. 

\begin{figure*}[t!]
    \centering
    \includegraphics[width=1\linewidth]{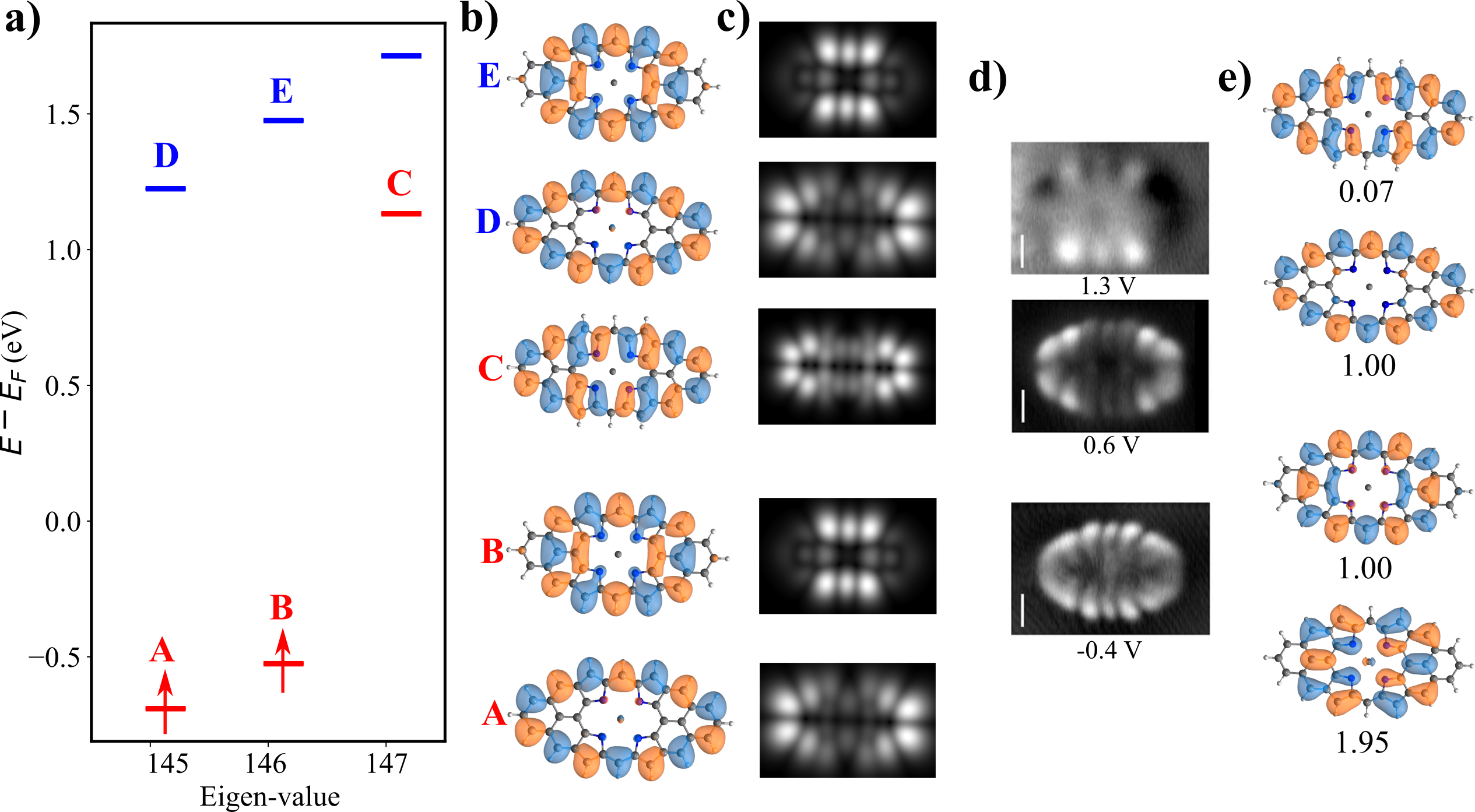}
    \caption{a) Spin polarized DFT-PBE0 eigen-value; b) canonical DFT orbitals obtained from spin polarized DFT-PBE0 calculations; c) simulated dI/dV maps of corresponding DFT-PBE0 orbitals; d) experimental dI/dV maps adapted with permission from ref. \cite{sun2020inducing} Copyright 2020 American Chemical Society; e) multi-reference natural orbitals obtained from CASCI(12,12) calculation with corresponding occupation for \textbf{APor$_2$} molecule.}
    \label{fig: dft_zn}
\end{figure*}

Next, we calculated Dyson orbitals $\varrho_{\pm},j$ using the multi-reference state to describe the transition from the neutral ground state $\Psi_0$ to charged states $\Psi_{\pm,j}$ driven by the tunneling process, as shown in Figure \ref{fig: TTATA_Dyson}a). The multi-reference electronic states are schematically represented by the expansion of the most important Slater determinants in the basis of close-shell DFT MOs. 
It is evident that the electronic structure of \textbf{TTAT} molecule is characterized by highly complex multi-reference states in both neutral and charged states, rendering the electronic properties and dI/dV maps challenging to elucidate via single-reference (DFT) methods.

Importantly, we observe that not all the transitions from the neutral quartet (S=3/2) ground state $\Psi_0$ are allowed. Namely, transitions to the singlet (S=0) states are prohibited, due to the fact that the single electron tunneling cannot change the total spin of the final state more than S=$\pm$1/2. The presence of the forbidden transition states strongly affects the resulting spectral function $\rho_s(\omega)$ and the emergence of ionic resonances in experimental STS spectra.   

Figures \ref{fig: TTATA_Dyson}b,c), present the multi-reference spectral function $\rho_s(\omega)$ and corresponding Dyson orbitals $\varrho_{\pm},j$ with relative transition strength. Dyson orbitals have a~notably different hierarchy compared to the canonical DFT MOs (Figure \ref{fig: TTATA_dft}b)). Importantly, neither natural orbitals obtained from the multi-reference CAS calculations,  see Figure \ref{fig: TTATA_dft}e), can explain the experimental STS data, which underlines the importance of the transition probabilities naturally included in the concept of Dyson orbitals. Comparing Figures \ref{fig: TTATA_Dyson}d-e) we find remarkable agreement between the simulated dI/dV maps derived from Dyson orbitals with the experimental dI/dV maps, validating the concept of multi-reference Dyson orbitals. 

\begin{figure*}[ht]
    \centering
    \includegraphics[width=1\linewidth]{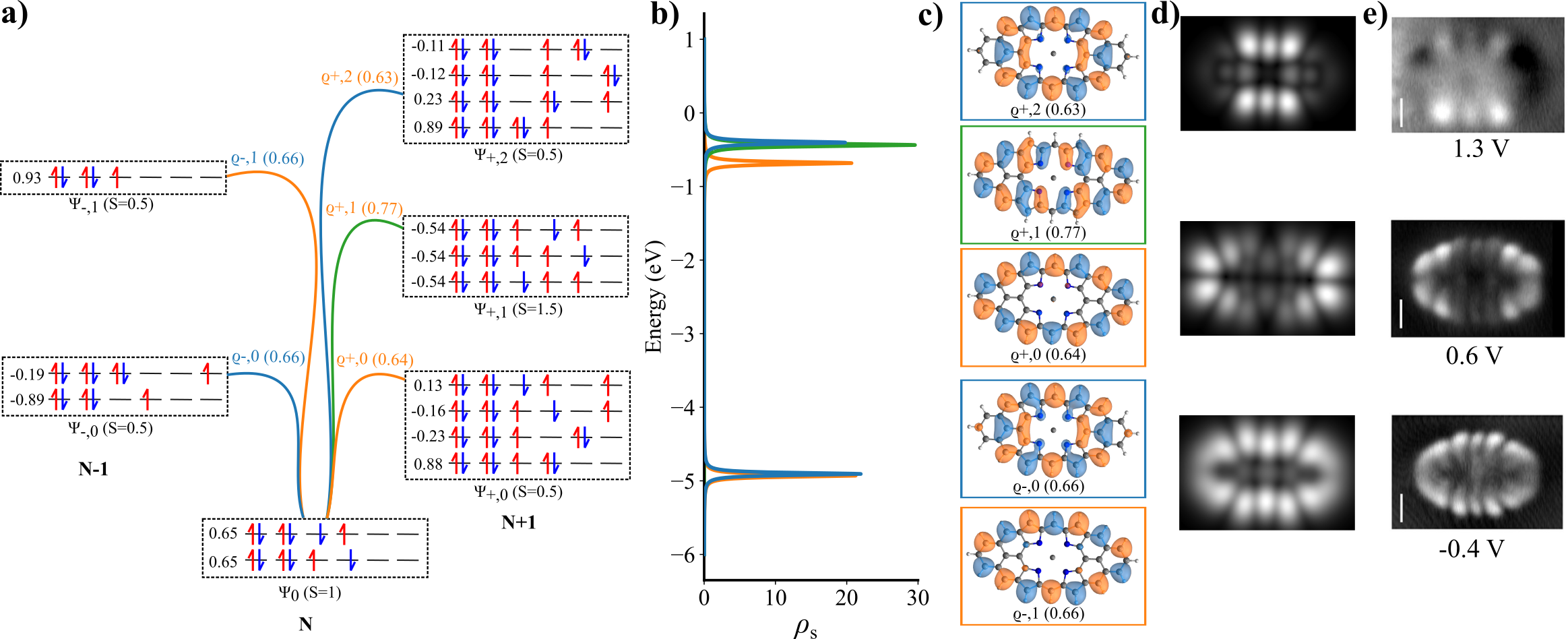}
    \caption{a) Multi-reference wavefunctions of neutral ground state and charge state multiplets; b) calculated multi-reference spectral function;  c) multi-reference Dyson orbitals obtained from CASCI(11,11) calculations with corresponding strengths; d) simulated dI/dV of corresponding Dyson orbitals using PP-STM code \cite{Krej2017} with CO-like tip, and dI/dV map correspoiding to $\rho-,0$ and $\rho-,1$ are added because they are very close in energy from the multi-reference spectral function ; e)  experimental dI/dV maps adapted with permission from ref. \cite{sun2020inducing} Copyright 2025 American Chemical Society for \textbf{APor$_2$} molecule. }
    \label{fig: Dyson_zn}
\end{figure*}

The second example consists of an open-shell Zn$^{\text{II}}$Porphyrin molecule (\textbf{APor$_2$}) presented by Sun et al. on Au(111) surface \cite{sun2020inducing}. The STS measurements reveal a~magnetic signal showing the coexistence of Kondo resonance as well as the spin-flip excitation signal at 19 meV. The origin of Kondo resonance as well as the spatial distribution of dI/dV maps were recently rationalized theoretically using the Kondo orbital concept \cite{calvo2024theoretical} considering the triplet ground state obtained from CAS calculations. Similarly, the spin excitation signal is theoretically rationalized by the triplet-singlet spin excitation \cite{calvo2024theoretical}.

Here, we will focus on explaining the STS measurements of ionic resonances observed at large bias voltages. As already pointed out in the original work \cite{sun2020inducing}, canonical DFT orbitals cannot fully explain the experimental STS data. Figure \ref{fig: dft_zn}a-b) shows calculated frontier canonical orbitals obtained from spin-polarized DFT calculation and corresponding dI/dV maps. We see that none of the SOMO states labeled A and B can reproduce the experimental dI/dV map of NIR obtained at -0.4 V. Similarly, the lowest PIR resonance acquired at 0.6 V does not match the simulated dI/dV map of SUMO state labeled C.     

According to the multi-reference CASCI(12,12) calculations performed using the DFT orbitals shown in Figure S4, the ground state wavefunctions \textbf{APor$_2$} in the neutral N and  charged N$\pm$1 states manifest a~strong multi-reference character, as shown in Figure \ref{fig: Dyson_zn}a). Therefore, it is essential to transcend the use of the canonical DFT orbitals and employ multi-reference Dyson orbitals to rationalize the experimental dI/dV data. 
In the multi-reference spectral function $\rho_s(\omega)$, shown in Figure \ref{fig: Dyson_zn}b), there are two peaks which are very close in energy with a~strong overlap. Combination of these two Dyson orbitals provides a very good match to the experimental dI/dV map of the first NIR acquired at -0.4 V, see Figures \ref{fig: Dyson_zn}d-e). Also, the first experimental dI/dV map of the first PIR agrees with the simulated dI/dV map corresponding to the Dyson orbital $\varrho_{+,0}$, which is attributed to the transition to the ground N+1 state $\Psi_{+,0}$.  The spatial resolution of the second PIR  fits well with the Dyson transition $\varrho_{+,2}$ to the second excited N+1 state $\Psi_{+,2}$ instead of the first excited state $\Psi_{+,1}$. Therefore, one would expect to observe first the Dyson transition $\varrho_{+,1}$ instead of $\varrho_{+,2}$. We tentatively attribute this deficiency in theory to the limited size of the active space used in the CASCI(12,12) calculations, as the energy difference between $\varrho_{+,1}$ and $\varrho_{+,2}$ decreases when increasing the active space. 

To summarize, the previous discussion clearly showed that STS measurements of strongly correlated molecules cannot provide reliable information about the canonical MOs, nor can they tell us the exact location of the electron before or after removal or addition. What STS measurements track are the transitions, well-represented by the concept of Dyson orbitals. While Dyson orbitals are also referred to as orbitals, they are distinct from the molecular orbitals we typically associate with molecules, as explained in \cite{truhlar2019orbitals}.
The association of dI/dV maps with one-electron MOs is problematic to say at least. Therefore, it should be carefully reconsidered when used to discuss general concepts such as the SHI effectx, as we will discuss in the next section.

\subsection{The SOMO/HOMO inversion effect and STS measurements}

\begin{figure*}[htb]
    \centering
    \includegraphics[width=1\linewidth]{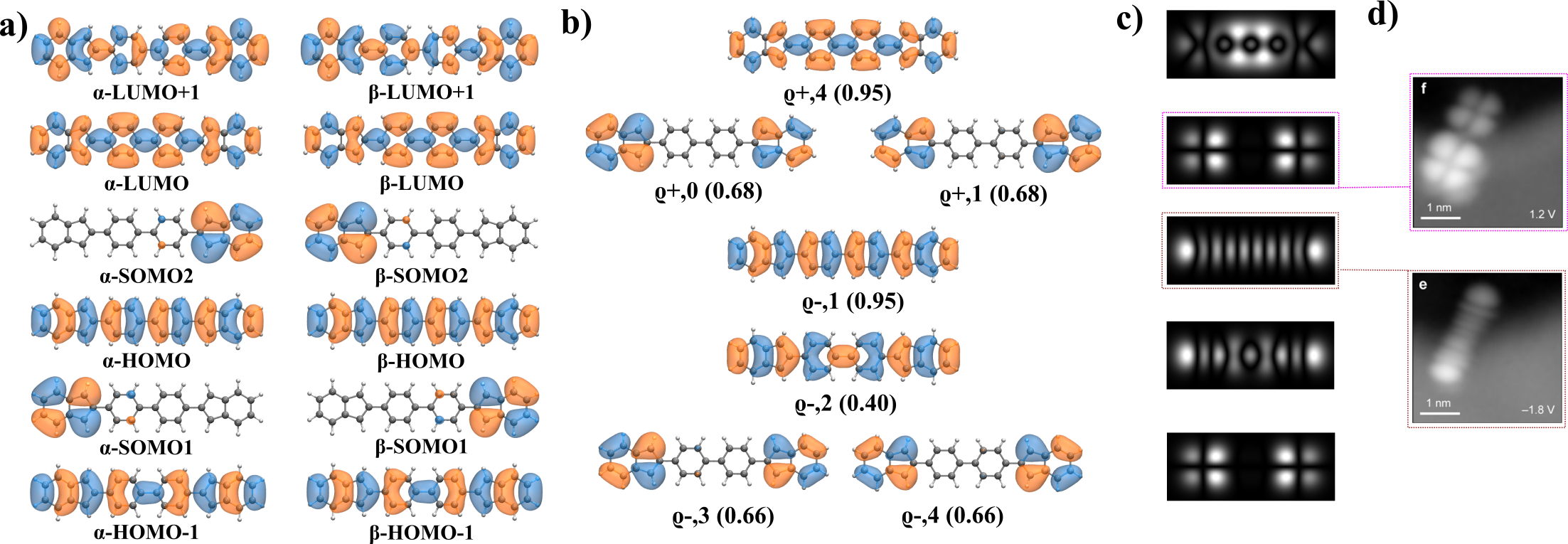}
    \caption{a) Canonical DFT orbitals obtained from spin polarized DFT-PBE0 calculations; b) multi-reference Dyson orbitals obtained from CASCI(12,12) calculations with corresponding strengths;  c) simulated dI/dV maps of corresponding Dyson orbitals using PP-STM code \cite{Krej2017} with CO-like tip;  d) experimental  constant current topography maps at indicated sample voltages, adapted with permission from ref. \cite{Mishra2024} Copyright 2024 American Chemical Society for \textbf{DNPAH} molecule.}
    \label{fig: dnpah_dft}
\end{figure*}

Recently, molecules that exhibit the SHI effect have received considerable attention both theoretically and experimentally \cite{Kumar2017,Kasemthaveechok2022,Westcott2000,Kusamoto2008}. These organic radicals feature an electron configuration in which the energy of the SOMO state is presumably below HOMO level. Thus, this concept is based on the assumption of an unusual arrangement of single-electron MOs that contradicts the traditional Aufbau principle of molecules. Recently published STM measurements \cite{Mishra2024} reported the presence of the SHI effect for a diradical nonbenzenoid open-shell polycyclic conjugated hydrocarbon (\textbf{DNPAH}). 

In that work, they found that low bias STM images matches well with the simulated dI/dV map corresponding to the canonical HOMO orbital obtained from DFT calculations. Moreover, the  spin-unrestricted DFT calculations give the canonical SOMO orbital below HOMO orbital in energy, as shown in Figure \ref{fig: dnpah_dft}a). They rationalized the agreement between the experimental and simulated dI/dV maps and the reverse energy ordering of the SOMO and HOMO canonical orbitals to the presence of the SHI effect.  However, given the previous discussion, there is the question whether STM measurements are in principle suitable to detect the order of MOs in strongly correlated polyradical molecules addressing the SHI effect. Similarly, one can ask what is the meaning of one-electron MOs in case of strongly correlated molecules. 

\begin{figure*}[t!]
    \centering
    \includegraphics[width=1\linewidth]{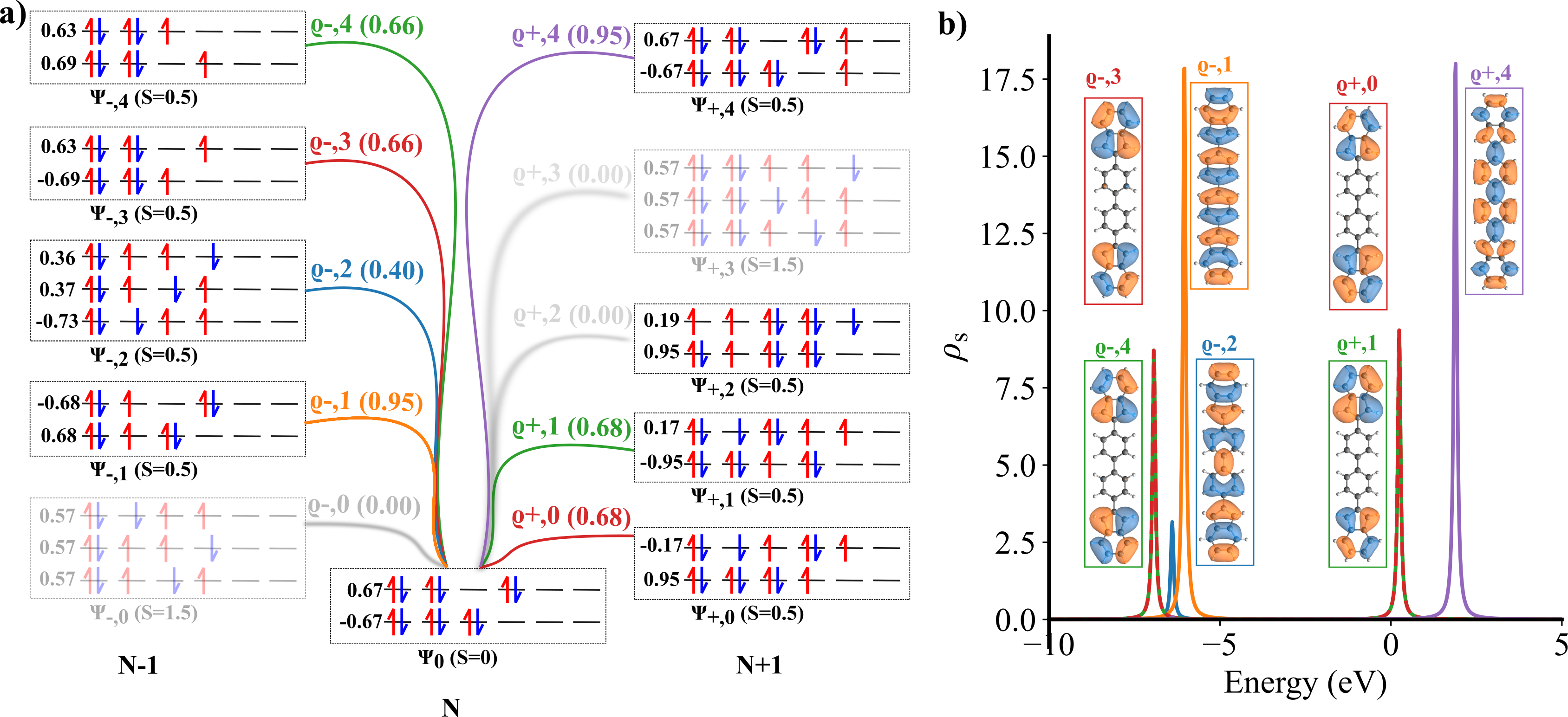}
    \caption{ a) Multi-reference wavefunctions of neutral ground state and charge state multiplet b) multi-reference spectral function for the \textbf{DNPAH} molecule.}
    \label{fig: dnpah-spectral_func}
\end{figure*}

To rationalize the STM measurements, we carried out multi-reference CASCI(12,12) calculations and Dyson orbitals. The DFT orbitals used for CAS calculations are shown in Figure S5. Note, that the lowest CAS Dyson orbitals show good agreement with the all-$\pi$ Dyson orbitals computed using the DMRG method, as presented in Figure S7.
The CAS calculation determines the diradical singlet ground N state $\Psi_0$ with two SONO orbitals with occupancy 1. Figure \ref{fig: dnpah-spectral_func}a) shows possible transitions from the neutral ground state $\Psi_0$ to the low-energy charged $\Psi_{\pm,j}$ states, including their transition strengths. We observe that two transitions from the singlet (S=0) ground state $\Psi_0$ to high spin charged states (S=1.5) $\Psi_{-,0}$ and $\Psi_{+,3}$, are not allowed due to the spin conservation. This can be understood by the fact that (de)attachment of a~single electron cannot convert the ground $\Psi_0$ state to those high spin $\Psi_{+,3}$ and $\Psi_{+,3}$ states, see Figure \ref{fig: dnpah-spectral_func}a).  
Figure \ref{fig: dnpah-spectral_func}b) shows the resulting spectral function with the allowed Dyson orbitals $\varrho_{\pm},j$. We can see that the simulated dI/dV map of the Dyson orbital $\varrho_{-,1}$ matches nicely to the experimental STM maps attributed to the lowest NIR. We can see, that the Dyson $\varrho_{-,1}$ orbital corresponds to the process, where an electron is detached mostly from HOMO orbital. This explains why simulated dI/dV maps obtained from the canonical HOMO orbital can approximate the experimental data.   
Similarly, the first PIR corresponds to an electron attachment causing the transition from the neutral ground $\Psi_0$ state to doubly degenerate $\Psi_{+,0}$ and $\Psi_{+,1}$ states. This process is represented by two similar Dyson orbitals  $\varrho_{+,0}$ and $\varrho_{+,1}$, see Figure \ref{fig: dnpah-spectral_func}b). The simulated dI/dV maps, see Figure \ref{fig: dnpah_dft}c), reproduces nicely the experimental STM maps in Figure \ref{fig: dnpah_dft}d). Here, the attachment process cannot be easily associated with single MOs, as the states have a strong multi-reference character.

From the discussion, it is evident that in particular transitions from the ground $N$ state to charged $N\pm1$ states due to the electron tunneling processes can significantly affect the resulting contrast of the dI/dV map. Therefore, the interpretation of STS measurements of polyradical molecules based on association with one-electron canonical MOs is problematic.

The primary source of confusion arises from comparing one-electron DFT canonical orbitals with STM/STS maps, where the energy ordering and orbital maps appear to align well with experimental observations. In a closed-shell molecule, we can define a unique energy ordering, often referred to as single-electron energy. However, for open-shell molecules, this concept loses its relevance, as a single energy cannot be unambiguously assigned to a specific MO.
As illustrated in Figure \ref{fig: pentacene-spectral_func}a,b), the spectral function exhibits distinct energy levels corresponding to each Dyson orbital, making it seemingly possible to assign an energy to these orbitals. However, for open-shell molecules, as shown in Figure \ref{fig: dnpah-spectral_func}b), the SOMO and SUMO do not have well-defined energy localization. Instead, their energies can be found both above the highest occupied molecular orbital (HOMO) and below another lower-energy orbital. This inconsistency makes it fundamentally contradictory to assign a single energy value to molecular orbitals derived from Dyson orbitals. 

Moreover, for the open-shell singlet ground state, the correct wavefunction is a linear combination of two Slater determinants corresponding to the configurations $ \vert \uparrow \downarrow \rangle + \vert \downarrow \uparrow  \rangle$, see Figure \ref{fig: dnpah-spectral_func}a). Single-determinant approaches, such as DFT method, inherently fail to capture this multi-reference character. Instead, they adopt one of the configurations, either $\vert \uparrow \downarrow \rangle$ or $\vert  \downarrow  \uparrow \rangle$, leading to a significant deviation from the actual physical state. To compensate for this, single-reference methods introduce an artificial spatial splitting of the frontier HOMO and LUMO orbitals, resulting in what is termed a \emph{broken-spin symmetry solution}. However, this solution lacks any physical meaning, especially regarding the energy ordering of canonical MOs. This artifact is particularly relevant for the correct interpretation of the SHI effect. Here, the energies of SOMO and HOMO states are simply a consequence of the inadequacies of the unrestricted DFT method rather than a genuine feature of the electronic structure. More rigorous analysis of the description of the open-shell singlet state is provided in supplementary material.

 The second problem is associated with the STS measurement itself, because they include transitions associated with the (de)attachment of electrons. Therefore, the STS/STM measurement itself influences the ground state $\Psi_0$ of the molecule. 
 Moreover, certain electronic transitions are inherently forbidden due to selection rules. As a result, STS spectra only capture allowed electronic transitions, meaning that STS measurements provide incomplete information about the electronic structure and energy ordering of molecular orbitals.The STS measurements provide information about the ionization resonances of molecules, which, especially in the case of open-shell molecules, can differ significantly from the MO picture. Thus, STM lacks the ability to assign energy ordering to single-electron molecular orbitals. Importantly, STM does not directly observe molecular orbitals or their intrinsic energy ordering, as previously highlighted in Ref. \cite{pham2017can}. STM measurements do not provide direct information about the electronic structure of a neutral molecule. Instead, they perturb the system and detect electronic transitions, which do not inherently reveal the molecular orbitals of the neutral state or their corresponding energy levels.

 If one constructs Dyson orbitals for \textbf{DNPAH} molecule based on closed-shell DFT orbitals, where the HOMO is lower in energy than the singly occupied state, the resulting Dyson orbital ordering remains consistent. This demonstrates that Dyson orbitals do not directly reflect the single-electron energies of molecular orbitals, but rather the transitions observed in STM.


\section{Conclusion}
In conclusion, we presented the concept of multi-reference Dyson orbitals, which provides a reliable methodology to rationalize STS/STM measurements. Moreover, the concept of Dyson orbitals also avoids controversy regarding the observation of MOs in STS experiments, which violates the basic principles of quantum mechanics.  It enables us to incorporate two important ingredients: i) the electronic transitions from the neutral ground state to charged low-energy states invoked by the single electron tunneling; ii) describing multi-reference electronic states of strongly correlated molecules. We showed that the concept of one-electron MOs can approximate STS spectra of close-shell molecules, while it provides a completely misleading picture in the case of strongly correlated molecules.  Therefore, the multi-reference Dyson orbitals are mandatory to rationalize STS measurements of strongly correlated polyradical molecules. We also demonstrate that estimating the energy order of single electron MOs from STS measurements of strongly correlated open-shell molecules is problematic. 

In this context, STS measurements for strongly correlated molecules cannot provide any information about energies of one-electron MOs. Therefore, STS measurements cannot address SOMO-HOMO inversion, which concept is inherently associated with one-electron approximation. 

Due to the ambiguity between single-electron MOs and STS measurements for the strongly correlated polyradical molecules, we recommend labeling individual resonances as negative/positive ionic resonances instead of the commonly adopted notation using canonical MOs, e.g. HOMO/LUMO, etc. We believe that changing the nomenclature will help to improve the somewhat short-sighted interpretations of STS/STM measurements.

\section{METHODS}
The geometries of all molecules were optimized in their respective ground states using density functional theory (DFT) as implemented in the FHI-AIMS software package \cite{blum2009ab}. The hybrid PBE0 functional \cite{Adamo1999_PBE0} was employed for these calculations, with the Tkatchenko-Scheffler \cite{tkatchenko2009accurate} method incorporated to account for van der Waals interactions. To accurately capture the electronic structure, Complete Active Space Configuration Interaction (CASCI) calculations were performed. The one- and two-body integrals were generated using the ORCA quantum chemistry software \cite{neese2012orca}, based on the closed-shell DFT orbitals obtained with the PBE0 \cite{Adamo1999_PBE0} exchange-correlation functional. For the pentacene, \textbf{APor$_2$}, and \textbf{DNPAH} molecules, an active space of CASCI(12,12) was employed. Due to the odd number of electrons in the \textbf{TTAT} molecule, the restricted open-shell Kohn-Sham (ROKS) approach was used, with an active space of CASCI(11,11). After constructing the full many-body Hamiltonian from these integrals, we diagonalized the Hamiltonian and constructed Dyson orbitals using our in-house many-body code. The simulated dI/dV maps of the Dyson orbitals were calculated using the Probe Particle Scanning Probe Microscopy (PP-SPM) code \cite{Krej2017} for a CO-like tip.


\section*{Acknowledgments}
The authors also thank Benjamin Lowe and Leo Gross for fruitful discussions. 
The author acknowledges support from the CzechNanoLab Research Infrastructure supported by MEYS CR (LM2023051) and the GACR project no. 23-05486S. M.K., M.L. and P.J. acknowledge the financial support from the grant CZ.02.01.01/00/22\_008/0004594 supported by MEYS.

\bibliography{bib.bib}
\end{document}


\oldmaketitle

\tableofcontents
\section{Non Uniqueness of Orbitals}\label{no_unique}
Besides the fact that molecular orbitals cannot be directly associated with any observable quantities, and they can be defined in multiple ways and STM does not provide information about the specific orbitals from which electrons are removed or where they go when added. Instead, STM offers spatial maps that capture transitions between the  $N$  and  $N \pm 1$  states, which are described by Dyson orbitals. As illustrated in Figure \ref{fig: benzene}, we can construct a many-body wavefunction using various types of orbitals like DFT orbitals, localized orbitals for example. The many-body natural orbitals and Dyson orbitals are uniquely defined, regardless of the specific orbitals chosen for the construction. In this sense, STM does not provide information about the molecular orbitals nor can it tell us the exact location of the electron before or after removal or addition. What STM can track are the transitions, which are well represented by Dyson orbitals. Although Dyson orbitals are also called orbitals, they are different from the molecular orbitals that we typically associate with molecules, as explained in \cite{truhlar2019orbitals}.

\begin{figure}[!htbp]
    \centering
    \includegraphics[width=1\linewidth]{Figures/benzene.png}
    \caption{ a) DFT orbitals b) many-body natural orbitals with the occupation below constructed by starting from the DFT orbitals. c) many-body Dyson orbitals constructed by starting from DFT orbitals.
    d) Local orbitals of $\pi$ space . e) many-body natural orbitals with occupation below constructed by starting from the the localized orbitals. f) many-body dyson orbitals constructed by starting from localized orbitals.}
    \label{fig: benzene}
\end{figure}

\section{Two electron in two orbitals}\label{Appendix BS_solution}
\textbf{Wavefunction Analysis of Diradical System in DFT Calculations}\\
Considering two electrons in two orbitals we can analyze the wavefunctions associated with diradical systems in Density Functional Theory (DFT) calculations. Here, we examine cases where the triplet and singlet configurations are the ground states, demonstrating the limitations of DFT in handling open-shell singlet states, especially when broken-symmetry solutions arise.

\textbf{Triplet State with $M_S = 1$}

Consider the case where we have two electrons with both spins up. The Slater determinant for the $M_S = +1$ triplet state can be expressed with two electrons in different orbitals as follows:

\[
\Psi_{S=1,M_{S}=+1} = \frac{1}{\sqrt{2}} \left|\begin{matrix}
\varphi_{A}(1)\alpha(1) & \varphi_{A}(2)\alpha(2) \\
\varphi_{B}(1)\alpha(1) & \varphi_{B}(2)\alpha(2)
\end{matrix}\right|
\]

Upon simplification:

\[
\Psi_{S=1,M_{S}=+1} = \frac{1}{\sqrt{2}} \left( \varphi_{A}(1)\alpha(1)\varphi_{B}(2)\alpha(2) - \varphi_{A}(2)\alpha(2)\varphi_{B}(1)\alpha(1) \right)
\]

Separating spin and spatial components, we get:

\[
\Psi_{S=1,M_{S}=+1} = \frac{1}{\sqrt{2}} \left( \varphi_{A}(1)\varphi_{B}(2) - \varphi_{A}(2)\varphi_{B}(1) \right) \alpha(1)\alpha(2)
\]

This wavefunction is antisymmetric in its total form, with the spatial part being antisymmetric and the spin part symmetric. It is an eigenfunction of the $\hat{S^2}$ operator:

\[
\hat{S^2} \Psi_{S=1} = S(S + 1) \Psi_{S=1} = 2 \Psi_{S=1}
\]

\textbf{Triplet State with $M_S = -1$}

For the $M_S = -1$ triplet state, the wavefunction is obtained by substituting $\alpha$ with $\beta$:

\[
\Psi_{S=1,M_{S}=-1} = \frac{1}{\sqrt{2}} \left( \varphi_{A}(1)\varphi_{B}(2) - \varphi_{A}(2)\varphi_{B}(1) \right) \beta(1)\beta(2)
\]

This state is also an eigenfunction of the $\hat{S^2}$ operator with the same eigenvalue.

\textbf{Singlet State with $M_S = 0$}

In the singlet $M_S = 0$ state, one electron is spin-up and the other is spin-down. The wavefunction, in determinant form, is:

\[
\Psi_{M_{S}=0} = \frac{1}{\sqrt{2}} \left|\begin{matrix}
\varphi_{A}(1)\alpha(1) & \varphi_{A}(2)\beta(2) \\
\varphi_{A}(1)\beta(1) & \varphi_{A}(2)\alpha(2)
\end{matrix}\right|
\]

After simplification, we have:

\[
\Psi_{M_{S}=0} = \frac{1}{\sqrt{2}} \left( \varphi_{A}(1)\alpha(1)\varphi_{A}(2)\beta(2) - \varphi_{A}(2)\alpha(2)\varphi_{A}(1)\beta(1) \right)
\]

This is a closed-shell singlet solution, with a symmetric spatial part and an antisymmetric spin part. It is also an eigenfunction of $\hat{S^2}$ with eigenvalue zero:

\[
\hat{S^2} \Psi_{S=0} = S(S + 1) \Psi_{S=0} = 0
\]

\textbf{Triplet State with $M_S = 0$: Broken-Symmetry Solution}

In the $M_S = 0$ triplet state, both electrons are in different orbitals with opposite spins. The broken-symmetry solutions for this state are as follows:

\begin{align*}
\Psi_{M_S=0} &= \frac{1}{\sqrt{2}} \left( \varphi_A(1)\varphi_B(2)\alpha(1)\beta(2) - \varphi_A(2)\varphi_B(1)\alpha(2)\beta(1) \right) \\
\Psi_{M_S=0} &= \frac{1}{\sqrt{2}} \left( \varphi_A(1)\varphi_B(2)\beta(1)\alpha(2) - \varphi_A(2)\varphi_B(1)\beta(2)\alpha(1) \right)
\end{align*}

Each broken-symmetry solution is not an eigenfunction of $\hat{S^2}$ and does not represent a pure spin state, resulting in spin contamination.

\textbf{Physical Interpretation of Broken-Symmetry Solutions}

In DFT, open-shell singlet solutions are approximated using broken-symmetry solutions, which are not eigenfunctions of $\hat{S^2}$. If we combine both broken-symmetry solutions, we obtain:

\[
\Psi_{S=1,M_{S}=0} = \frac{1}{2} \left( \varphi_{A}(1)\varphi_{B}(2) - \varphi_{A}(2)\varphi_{B}(1) \right) \left( \alpha(1)\beta(2) + \beta(1)\alpha(2) \right)
\]

This is the triplet state with $M_S = 0$. By contrast, if we subtract these solutions, we get the singlet state:

\[
\Psi_{S=0,M_{S}=0} = \frac{1}{2} \left( \varphi_{A}(1)\varphi_{B}(2) + \varphi_{A}(2)\varphi_{B}(1) \right) \left( \alpha(1)\beta(2) - \beta(1)\alpha(2) \right)
\]

Here, the spatial part is symmetric, and the spin part is antisymmetric, representing the singlet state.

\textbf{Checking for Broken-Symmetry Solutions}

To verify if a DFT solution is a broken-symmetry solution, we calculate $\langle S^2 \rangle$ as \cite{andrews1991spin} :

\[
\langle S^2 \rangle_{MFH} = \langle S^2 \rangle_{exact} + N_{\beta} - \sum_{ij}^{occ} |\langle \psi^{\alpha}_{i}|\psi^{\beta}_{j} \rangle|^{2}
\]

where:

\[
\langle S^2 \rangle_{exact} = \left( \frac{N_{\alpha} - N_{\beta}}{2} \right) \left( \frac{N_{\alpha} - N_{\beta}}{2} + 1 \right)
\]

Broken-symmetry solutions are commonly used in DFT but lack pure spin character due to spin contamination, impacting their utility in describing accurate electronic structure.


\section{Recovering molecular orbitals}

In this section, we show explicitly how in the limit where a one-electron approximation holds, the Dyson orbitals are reproduced by the canonical orbitals. Let us consider two states, $\vert \Psi_0 \rangle$, $\vert \Psi_+ \rangle$, respectively for the neutral state (with $N$ electrons) and the charged system ($N+1$ electrons).

Let us assume these states are well described by Slater determinants, as is the case in one-electron methods like DFT:

$$ \vert \Psi_+ \rangle = \text{Det}\left( \phi_1,\ldots,\phi_{N+1}  \right) = \dfrac{1}{\sqrt{(N+1)!}} \sum_{\sigma \in S_{N+1}} (-1)^{\vert \sigma \vert } \prod_{j=1}^{N+1} \phi_{\sigma(j)}(x_j)$$
$$\vert \Psi_0 \rangle = \text{Det}\left( \varphi_1,\ldots,\varphi_{N}  \right) = \dfrac{1}{\sqrt{N!}} \sum_{\tau \in S_{N}} (-1)^{\vert \tau \vert } \prod_{j=1}^{N} \varphi_{\tau(j)}(x_j),$$

where in general we have different spin-orbitals $\{ \phi \}$ and $\{ \varphi \}$ (which do not need to be mutually orthogonal) to describe the two wavefunctions.

The Dyson orbital corresponding to these two states is given by the multireference overlap:
$$\varphi_{+}(x_{N+1}) = \sqrt{N+1}\int \text{Det} \left( \phi_1,\ldots,\phi_{N+1}   \right) \text{Det} \left( \varphi_1,\ldots,\varphi_N   \right) dx_1 \ldots dx_N. $$

Now let us expand the spin-orbitals in a common basis, $\{ \chi_\mu(x) \},$ which we can think of as the basis set of our one-electron calculation:
$$\phi_a(x) = \sum_\mu \phi_{a\mu} \chi_\mu(x) $$
and
$$\varphi_b(x) = \sum_\nu \varphi_{b\nu} \chi_\nu(x). $$
Such expansion is the result of performing DFT or any other one-electron method for the neutral and charged system.

Let us further define the overlap matrix $\boldsymbol{D}$ with elements $D_{a,b}$:
$$D_{a,b} = \int \phi_a(x) \varphi_b(x) dx.$$

Notice that if the orbitals for the neutral and charged system are the same (as is usually assumed), the matrix $\boldsymbol{D}$ is the identity.

In general, however, by defining the basis set overlap $S_{\mu \nu} = \int \chi_\mu (x) \chi_\nu(x) dx$, we see that we can compactly write:
$$\boldsymbol{D} = \boldsymbol{\phi} \boldsymbol{S} \boldsymbol{\varphi}^t$$.
 
Introducing further the notation
$$d(\sigma) = \dfrac{(-1)^{\vert \sigma \vert}}{N!} \sum_{\tau \in S_N} (-1)^{\vert \tau \vert} \prod_{j=1}^N D_{\sigma(j) \tau(j)},$$
a straight-forward manipulation yields the expression:
$$\varphi_+(x) = \sum_{\sigma \in S_{N+1}} d(\sigma) \phi_{\sigma(N+1)}(x).$$

The matrix $\boldsymbol{D}$ will typically be close to the identity. In most practical calculations, the shape of the orbitals is not reoptimized for the charged system. In such cases, $\boldsymbol{D}$ is \textit{exactly} the identity, since $d(\sigma)$ is 1 if $\sigma(N+1) = N+1$ and $d(\sigma) = 0$ otherwise, because in this case $D_{\sigma(j) \tau(j)} = \delta_{\sigma(j) \tau(j)}.$ 

In such a case is then clear that:
$$\varphi_+(x) = \phi_{N+1}(x),$$
so that the Dyson orbital coincides exactly with the canonical molecular orbital.
\\

This simple calculation shows how the molecular orbitals serve as a first approximation to the multireference transitions. However, even without treating the correlation beyond the one-electron picture, we see how just a calculation of molecular orbitals for a neutral system will fail to yield the correct ordering of the observed transitions, since it also does not account in general for the re-optimization of the electronic structure for different charged states.
\\

\section{DFT Orbitals used for CASCI}

\begin{figure}[!htbp]
    \centering
    \includegraphics[width=1\linewidth]{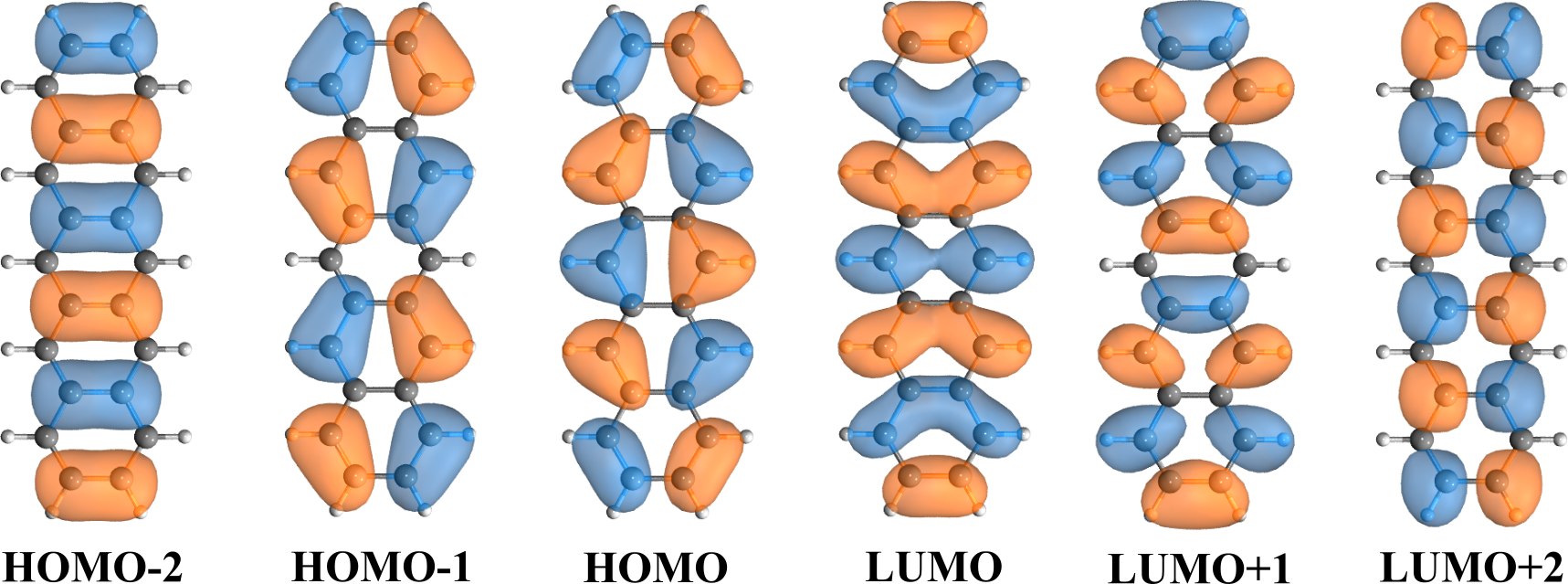}
    \caption{ Frontier orbitals around the Fermi level used in CAS(12,12) obtained from the spin-unpolarized DFT-PBE0 calculation for the pentacene molecule.}
    \label{fig: pentacene_cl_dft}
\end{figure}

\begin{figure}[!htbp]
    \centering
    \includegraphics[width=1\linewidth]{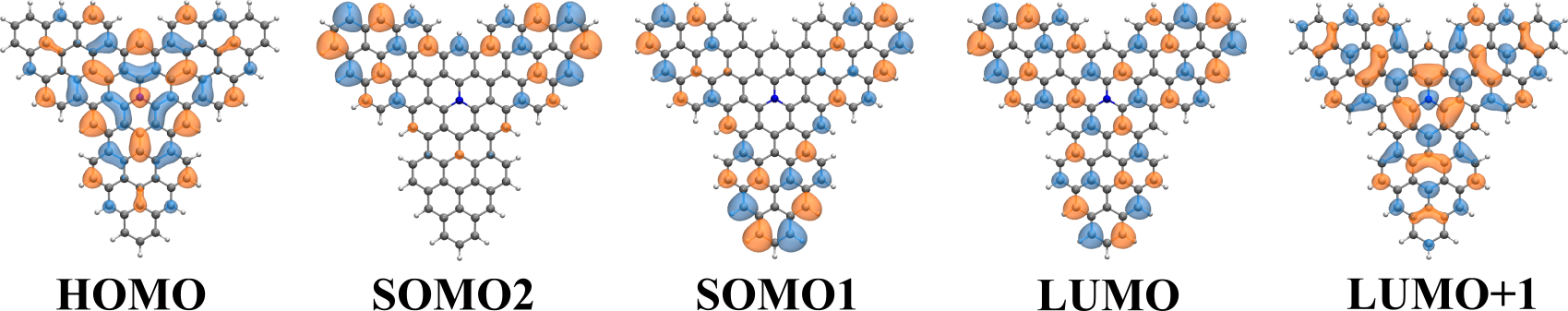}
    \caption{ Frontier orbitals around the Fermi level used in CAS(11,11) obtained from the  restricted open-shell Kohn-Sham (ROKS) DFT-PBE0 calculation for the \textbf{TTAT} molecule.}
    \label{fig: TTATA_cl_dft}
\end{figure}

\begin{figure}[!htbp]
    \centering
    \includegraphics[width=1\linewidth]{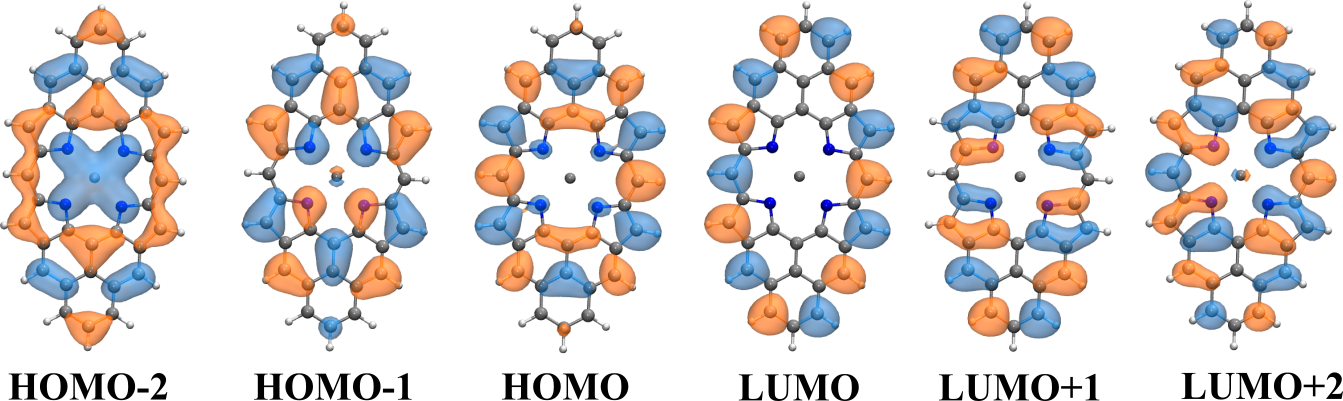}
    \caption{ Frontier orbitals around the Fermi level used in CAS(12,12) obtained from the spin-unpolarized DFT-PBE0 calculation for the \textbf{APor$_2$} molecule.}
    \label{fig: zn_cl_dft}
\end{figure}

\begin{figure}[!htbp]
    \centering
    \includegraphics[width=1\linewidth]{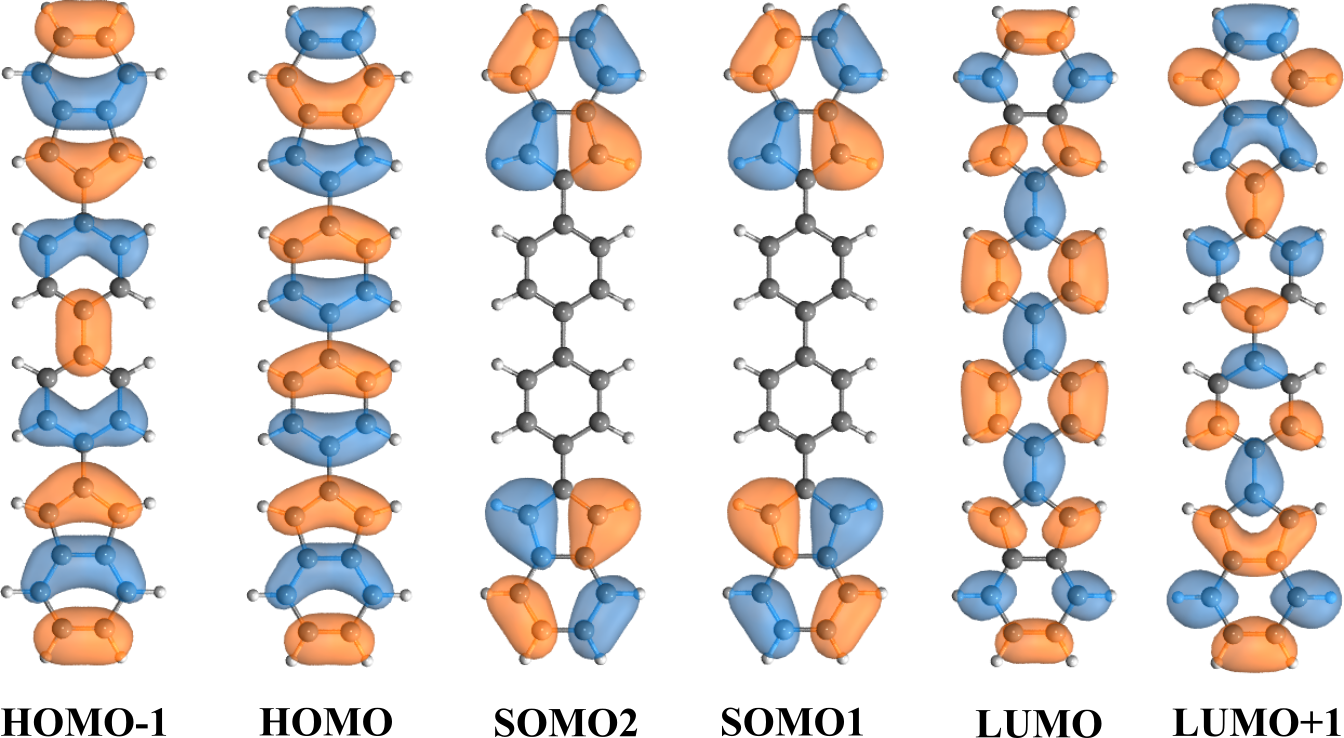}
    \caption{ Frontier orbitals around the Fermi level used in CAS(12,12) obtained from the spin-unpolarized DFT-PBE0 calculation the \textbf{DNPAH} molecule.}
    \label{fig: DNPAH_cl_dft}
\end{figure}

\section{Wavefunction Representation of the \textbf{TTAT} Molecule in Different  $S_z$  subspace}
The wavefunction of the \textbf{TTAT} molecule in its neutral ground state can be represented in the  $S_z = 1.5$  spin projection where it can be represented in a single Slater determinant. For charged states, the wavefunction can be described using an  $S_z = 1.0$  representation. Importantly, the Dyson orbitals remain invariant with respect to the choice of  $S_z$ , indicating that the fundamental electronic transitions are independent of the spin projection selected for the description.

However, it is crucial to recognize that within the charge-state multiplets, there exist additional electronic states with  $S_z = 0$  (corresponding to open-shell singlet configurations). These states are not included when constraining the Hamiltonian to a higher spin representation. Consequently, transitions involving single-electron processes that would typically manifest in Dyson orbitals become inaccessible due to the restriction imposed on the Hamiltonian. Specifically, any state with a total spin projection lower than the selected  $S_z$  of the Hamiltonian does not appear in the computed energy spectrum, as the Hamiltonian inherently excludes lower-spin configurations under these constraints.

\begin{figure}[!htbp]
    \centering
    \includegraphics[width=1\linewidth]{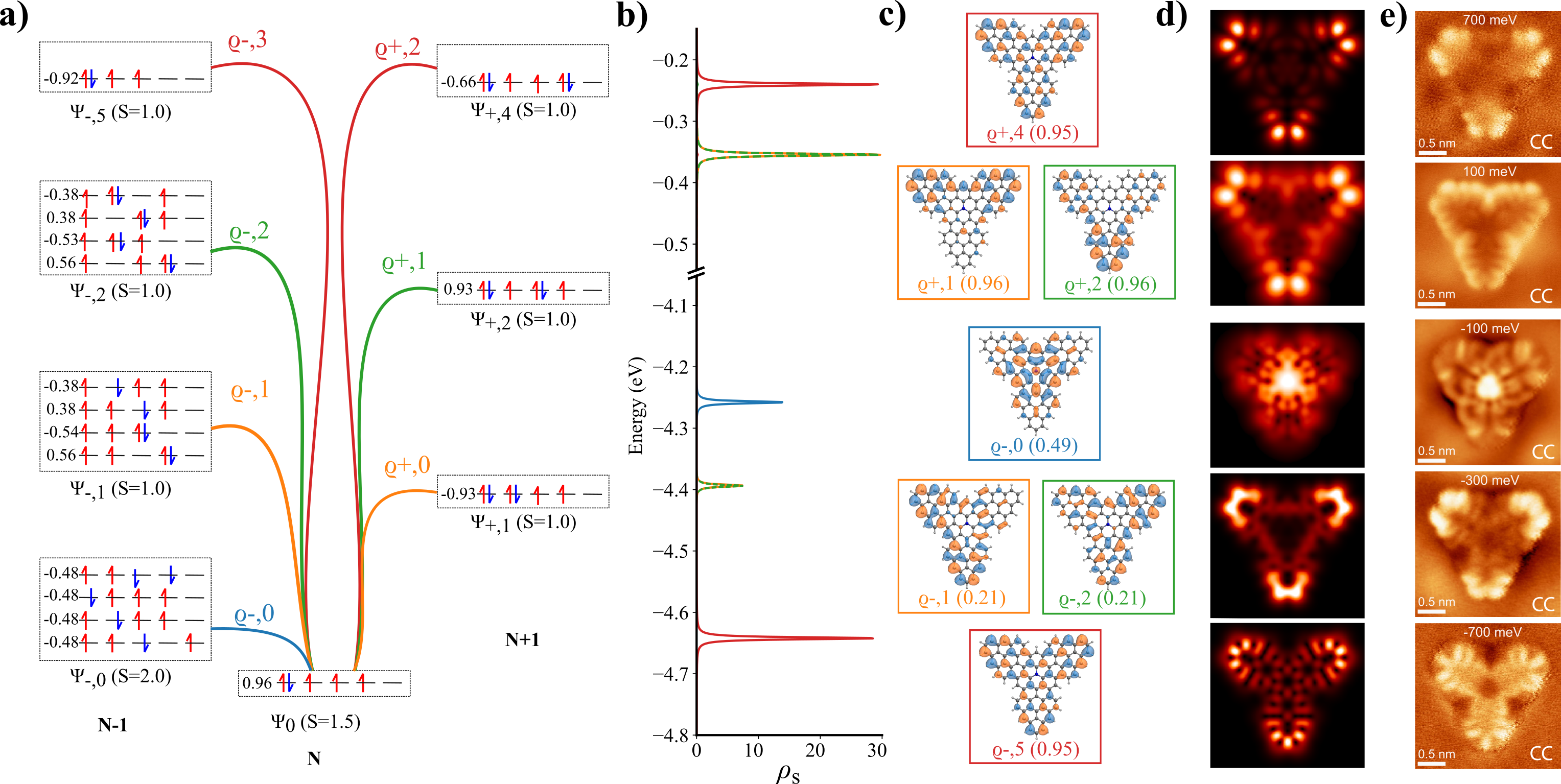}
    \caption{ a) Multi-reference wavefunctions of neutral ground state (in $S_z=1.5$ subspace) and charge state multiplets (in $S_z=1.0$ subspace); b) calculated multi-reference spectral function;  c) multi-reference Dyson orbitals obtained from CAS(11,11) calculations with corresponding strengths; d) simulated dI/dV of corresponding Dyson orbitals using PP-STM code \cite{Krej2017} with CO-like tip; e)  experimental dI/dV maps adapted with permission from ref. \cite{} Copyright 2025 American Chemical Society for \textbf{TTAT} molecule.  }
    \label{fig: TTATA_Dyson}
\end{figure}

\clearpage
\section{DMRG Dyson Orbitals of DNPAH diradical}

\begin{figure}[!htbp]
    \centering
    \includegraphics[width=1\linewidth]{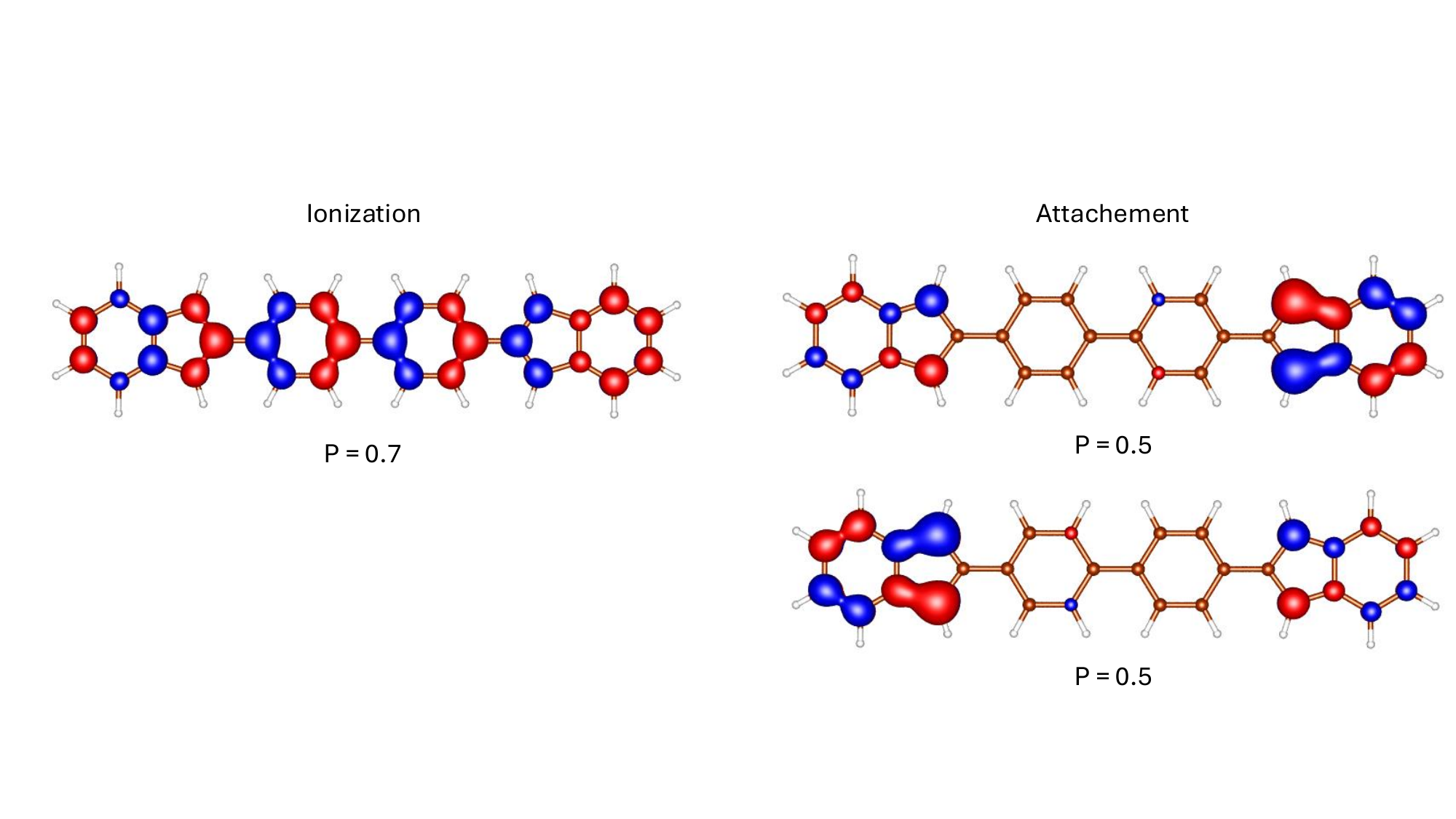}
    \caption{All-$\pi$ Dyson orbitals (along with their probability factors) corresponding to the lowest-energy ionization and electron attachment processes were computed using the DMRG method. Note that the electron attachment Dyson orbitals are degenerate and therefore combine in the final STM signal.}
    \label{fig: dmrg_dyson}
\end{figure}

We have also performed all-$\pi$ CAS calculations to assess the quality of the lowest CAS(12,12) Dyson orbitals. Specifically, we carried out DMRG-SCF(30,30)/cc-pVDZ orbital optimization with a fixed DMRG bond dimension of $M = 1000$. Subsequently, we performed high-accuracy DMRG calculations for both the neutral and charged species, employing dynamical block state selection (DBSS) \cite{legeza_2003a} with a predefined truncation error of $10^{-6}$. The resulting Dyson orbitals are presented in Figure \ref{fig: dmrg_dyson}. Orbital optimization was conducted using ORCA \cite{neese2012orca}, while the DMRG calculations were performed with the MOLMPS program \cite{Brabec2020}.

\clearpage

\bibliography{bib.bib}